\documentclass{ieeeaccess}

\usepackage[english]{babel}
\usepackage{url}
\usepackage{ifthen}
\usepackage{cite}
\usepackage[cmex10]{amsmath} 
\usepackage{longtable}
\usepackage{makecell}
\usepackage{algorithm,algorithmic}
                             
\usepackage{amssymb}
\usepackage{amsfonts}
\usepackage{xifthen,xparse}
\usepackage{float}
\usepackage{caption}

\usepackage{tikz,pgfplots,tkz-graph,color}
\pgfplotsset{compat=newest}
\pgfplotsset{plot coordinates/math parser=false}
\usetikzlibrary{decorations.markings,calc,positioning,matrix}
\usetikzlibrary{quantikz}

\NewSpotColorSpace{PANTONE}
  \AddSpotColor{PANTONE} {PANTONE3015C} {PANTONE\SpotSpace 3015\SpotSpace C} {1 0.3 0 0.2}
  \SetPageColorSpace{PANTONE}%
    
\usepackage{graphicx}

\usepackage{textcomp}

\allowdisplaybreaks

\newif\iffull
\fulltrue

\usepackage{listings}

\definecolor{dkgreen}{rgb}{0,0.6,0}
\definecolor{gray}{rgb}{0.5,0.5,0.5}
\definecolor{mauve}{rgb}{0.58,0,0.82}

\definecolor{myblue}{RGB}{0,0,255}
\definecolor{mygreen}{RGB}{28,172,0}
\definecolor{mylilas}{RGB}{170,55,241}
\definecolor{bitcolor}{rgb}{1,0.84314,0}
\definecolor{checkcolor}{rgb}{0.52941,0.80784,1}

\newtheorem{theorem}{Theorem}
\newtheorem{corollary}[theorem]{Corollary}
\newtheorem{remark}[theorem]{Remark}

\newtheorem{definition}[theorem]{Definition}

\newtheorem{fact}[theorem]{Fact}

\NewDocumentCommand\ketbra{+m+g}{%
  \IfNoValueTF{#2}
    {\left\lvert #1 \right\rangle \left\langle #1 \right\vert}
  {\left\lvert #1 \right\rangle \left\langle #2 \right\rvert}%
}

\newcommand{\MCC}{\mathcal{C}}

\newcommand{\vecnot}[1]{\underline{#1}}
\newcommand{\bg}{\boldsymbol{g}}
\newcommand{\llbr}{[\![}
\newcommand{\rrbr}{]\!]}
\newcommand{\cnot}[2]{\text{CNOT}_{#1 \rightarrow #2}}
\newcommand{\lcnot}[2]{\overline{\text{CNOT}}_{#1 \rightarrow #2}}
\newcommand{\cz}[2]{\text{CZ}_{#1#2}}
\newcommand{\lcz}[2]{\overline{\text{CZ}}_{#1#2}}
\newcommand{\lX}{\bar{X}}
\newcommand{\lZ}{\bar{Z}}

\newcommand{\syminn}[2]{\langle #1, #2 \rangle_{\text{s}}}

\def\BibTeX{{\rm B\kern-.05em{\sc i\kern-.025em b}\kern-.08em
    T\kern-.1667em\lower.7ex\hbox{E}\kern-.125emX}}    
    
\let\emph\textit

\begin{document}
\history{Date of publication xxxx 00, 0000, date of current version xxxx 00, 0000.}
\doi{}

\title{Logical Clifford Synthesis for Stabilizer Codes}
\author{\uppercase{Narayanan Rengaswamy}\authorrefmark{1}, \IEEEmembership{Member, IEEE},
\uppercase{Robert Calderbank}\authorrefmark{1}, \IEEEmembership{Fellow, IEEE}, \uppercase{Swanand Kadhe}\authorrefmark{2}, \IEEEmembership{Member, IEEE}, and \uppercase{Henry D. Pfister}\authorrefmark{1}, \IEEEmembership{Senior Member, IEEE}}
\address[1]{Department of Electrical and Computer Engineering, Duke University, Durham, NC 27708 USA (e-mail: \{ narayanan.rengaswamy, robert.calderbank, henry.pfister \}@duke.edu)}
\address[2]{Department of Electrical Engineering and Computer Sciences, University of California, Berkeley, CA 94720 USA (e-mail: swanand.kadhe@berkeley.edu)}
\tfootnote{Part of this work has been presented at the 2018 IEEE International Symposium on Information Theory~\cite{Rengaswamy-isit18}. This work was supported in part by the National Science Foundation (NSF) under Grant Nos. 1718494, 1908730 and 1910571. Any opinions, findings, conclusions, and recommendations expressed in this material are those of the authors and do not necessarily reflect the views of these sponsors.}

\markboth
{N. Rengaswamy \headeretal: Logical Clifford Synthesis for Stabilizer Codes}
{N. Rengaswamy \headeretal: Logical Clifford Synthesis for Stabilizer Codes}

\corresp{Corresponding author: Narayanan Rengaswamy (e-mail: narayanan.rengaswamy@duke.edu).}

\begin{abstract}
Quantum error-correcting codes are used to protect qubits involved in quantum computation. 
This process requires logical operators to be translated into physical operators acting on physical quantum states. 
We propose a mathematical framework for synthesizing physical circuits that implement logical Clifford operators for stabilizer codes. 
Circuit synthesis is enabled by representing the desired physical Clifford operator in $\mathbb{C}^{N \times N}$ as a $2m \times 2m$ binary symplectic matrix, where $N = 2^m$. 
We prove two theorems that use symplectic transvections to efficiently enumerate all binary symplectic matrices that satisfy a system of linear equations.
As a corollary, we prove that for an $\llbr m,k \rrbr$ stabilizer code every logical Clifford operator has $2^{r(r+1)/2}$ symplectic solutions, where $r = m-k$, up to stabilizer degeneracy. 
The desired physical circuits are then obtained by decomposing each solution into a product of elementary symplectic matrices, that correspond to elementary circuits. 
This enumeration of all physical realizations enables optimization over the ensemble with respect to a suitable metric.
Furthermore, we show that any circuit that normalizes the stabilizer can be transformed into a circuit that centralizes the stabilizer, while realizing the same logical operation. 
Our method of circuit synthesis can be applied to any stabilizer code, and this paper discusses a proof of concept synthesis for the $\llbr 6,4,2 \rrbr$ CSS code.
Programs implementing the algorithms in this paper, which includes routines to solve for binary symplectic solutions of general linear systems and our overall LCS (logical circuit synthesis) algorithm, can be found at \url{https://github.com/nrenga/symplectic-arxiv18a}.
\end{abstract}


\titlepgskip=-15pt

\maketitle

\section{Introduction}
\label{sec:intro}

\PARstart{I}{t} is expected that universal fault-tolerant quantum computation will be achieved by employing quantum error-correcting codes (QECCs) to protect the information stored in the quantum computer and to enable error-resilient computation on that data.
The first QECC was discovered by Shor~\cite{Shor-physreva95}, and subsequently, a systematic framework was developed by Calderbank, Shor and Steane~\cite{Calderbank-physreva96,Steane-physreva96} to translate (pairs of) classical error-correcting codes into QECCs. 
Codes produced using this framework are referred to as \emph{CSS} codes.
The general class of \emph{stabilizer} codes includes CSS codes as a special case and was introduced by Calderbank, Rains, Shor and Sloane~\cite{Calderbank-it98*2}, and by Gottesman~\cite{Gottesman-phd97}.
These codes, and their variations~\cite{Bacon-pra06,Yoder-arxiv17}, still remain the preferred class of codes for realizing error-resilient quantum computation in practice.

The \emph{Clifford hierarchy} of unitary operators was defined to help demonstrate that universal quantum computation can be realized via the teleportation protocol~\cite{Gottesman-nature99}.
The first level $\MCC^{(1)}$ in the hierarchy is the Pauli group of unitary operators, and subsequent levels $\MCC^{(\ell)}, \ell \geq 2$, are defined recursively as those unitary operators that map the Pauli group into $\MCC^{(\ell-1)}$, under conjugation.
By this definition, the second level is the normalizer of the Pauli group in the unitary group, and hence $\MCC^{(2)}$ is the Clifford group~\cite{Calderbank-it98*2}.
It is well-known that the levels $\MCC^{(\ell)}$ do not form a group for $\ell \geq 3$, but that the Clifford group along with \emph{any} unitary in $\MCC^{(3)}$ can be used to approximate an arbitrary unitary operator up to any desired precision.
(Note that using a simple inductive argument it can be proven that each level in the hierarchy is closed under multiplication by Clifford group elements.)
Therefore, the standard strategy for realizing universal computation with QECCs is to first synthesize\footnote{By ``synthesize'' we mean determine the logical operator, i.e., a circuit on the physical qubits of the QECC, that realizes the action of the given unitary operator on the logical qubits of that QECC.} logical Paulis, then logical Cliffords, and finally some logical non-Clifford in the third level of the Clifford hierarchy.
In this paper, we will be primarily concerned with logical Cliffords because specific QECCs, such as \emph{tri-orthogonal codes}~\cite{Bravyi-pra12}, can be used to distill \emph{magic states}~\cite{Bravyi-pra05} for a non-Clifford gate in $\MCC^{(3)}$, and these states can then be ``injected'' into the computation via teleportation in order to realize the action of that gate at the logical level~\cite{Gottesman-nature99}.
Hence, any circuit implemented on the computer equipped with error-correction might be expected to consist only of Clifford gates, augmented with ancilla magic states, and Pauli measurements.

For the task of synthesizing the logical Pauli operators for stabilizer codes, the first algorithm was introduced by Gottesman~\cite[Sec.~4]{Gottesman-phd97} and subsequently, another algorithm based on a symplectic Gram-Schmidt procedure was proposed by Wilde~\cite{Wilde-physreva09}. 
The latter is closely related to earlier work by Brun et al.~\cite{Brun-science06,Brun-it14}.
Since the logical Paulis are inputs to our algorithm that synthesizes logical Clifford operators for stabilizer codes, we will consider the above two procedures to be ``preprocessors'' for our algorithm.

Given the logical Pauli operators for an $\llbr m,k \rrbr$ stabilizer QECC, that encodes $k$ logical qubits into $m$ physical qubits, physical Clifford realizations of Clifford operators on the logical qubits can be represented by $2m \times 2m$ binary symplectic matrices, thereby reducing the complexity \emph{dramatically} from $2^{2m}$ complex variables to $4m^2$ binary variables (see~\cite{Calderbank-physrevlett97,Gottesman-arxiv09} and Section~\ref{sec:operators}).
We exploit this fact to propose an algorithm that efficiently assembles \emph{all} $2^{r(r+1)/2}$, where $r = m-k$, symplectic matrices representing physical Clifford operators (circuits) that realize a given logical Clifford operator on the protected qubits.
We will refer to this procedure as the \emph{Logical Clifford Synthesis (LCS) algorithm}.
Here, each symplectic solution represents an equivalence class of Clifford circuits, all of which ``propagate'' input Pauli operators through them in an identical fashion (see Section~\ref{sec:logical_clifford_synthesis}).
Moreover, as we will discuss later in the context of the algorithm, the other degrees of freedom not captured by our algorithm are those provided by stabilizers (see Remark~\ref{rem:stab_freedom}). 
But, at the cost of some increased computational complexity, the algorithm can easily be modified to account for these stabilizer degrees of freedom.
Hence, our work makes it possible to optimize the choice of circuit with respect to a suitable metric, that might be a function of the quantum hardware.
Note that our approach here is to determine \emph{unitary} physical operations to realize a specific logical (Clifford) operation, and this is distinct from operations such as lattice surgery~\cite{Litinski-quantum19} that are used to perform logical operations on topological codes.


The primary contributions of this paper are the four theorems that we state and prove in Section~\ref{sec:lcs_algorithm}, and the main LCS algorithm (Algorithm~\ref{alg:log_ops}) which builds on the results of these theorems.
These results form part of a larger program for fault-tolerant quantum computation, where the goal is to achieve reliability by using classical computers to track and control physical quantum systems, and perform error correction only as needed.

We note that there are several works that focus on exactly decomposing, or approximating, an arbitrary unitary operator as a sequence of operators from a fixed \emph{instruction set}, such as Clifford + $T$~\cite{Kliuchnikov-prl13,Amy-tcad13,Maslov-it17,Fagan-eptcs19,Iten-arxiv19,Duncan-arxiv19}.
However, these works do not consider the problem of circuit synthesis or optimization over different realizations of unitary operators on the \emph{encoded} space.
We also note that there exists several works in the literature that study this problem for specific codes and operations, e.g., see~\cite{Gottesman-phd97,Bacon-pra06,Fowler-arxiv12,Grassl-isit13,Kubica-pra15,Yoder-arxiv17,Chao-arxiv17b}. 
However, we believe our work is the first to propose a systematic framework to address this problem for general stabilizer codes, and hence enable automated circuit synthesis for encoded Clifford operators.
This procedure is more systematic in considering all degrees of freedom than conjugating the desired logical operator by the encoding circuit for the QECC.

Recently, we have used the LCS algorithm to translate the unitary $2$-design we constructed from classical Kerdock codes into a \emph{logical} unitary $2$-design~\cite{Can-arxiv19}, and in general any design consisting of only Clifford elements can be transformed into a logical design using our algorithm. 
An implementation of the design is available at: \url{https://github.com/nrenga/symplectic-arxiv18a}.
This finds direct application in the \emph{logical randomized benchmarking} protocol proposed by Combes et al.~\cite{Combes-arxiv17}.
This protocol is a more robust procedure to estimate logical gate fidelities than extrapolating results from randomized benchmarking performed on physical gates~\cite{Magesan-physreva12}.
Now we discuss some more motivations and potential applications for the LCS algorithm.


\subsection{Noise Variation in Quantum Systems}

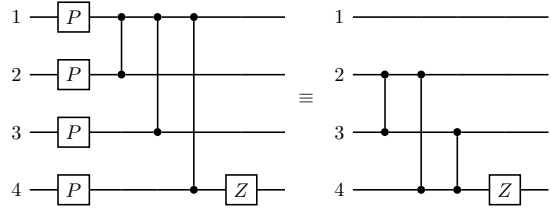
\begin{figure}
\begin{center}

\scalebox{0.75}{%
\begin{tikzcd}
\lstick{$1$} & \gate{P} & \ctrl{1} & \ctrl{2} & \ctrl{3} & \qw & \qw \\
\lstick{$2$} & \gate{P} & \control{} & \qw & \qw & \qw & \qw \\
\lstick{$3$} & \gate{P} & \qw & \control{} & \qw & \qw & \qw \\
\lstick{$4$} & \gate{P} & \qw & \qw & \control{} & \gate{Z} & \qw
\end{tikzcd}
$\equiv$
\begin{tikzcd}
\lstick{$1$} & \qw & \qw & \qw & \ghost{Z}\qw & \qw \\
\lstick{$2$} & \ctrl{1} & \ctrl{2} & \qw & \ghost{Z}\qw & \qw \\
\lstick{$3$} & \control{} & \qw & \ctrl{1} & \ghost{Z}\qw & \qw \\
\lstick{$4$} & \qw & \control{} & \control{} & \gate{Z} & \qw 
\end{tikzcd}
}

\caption{\label{fig:cz_412}Two physical circuits that realize the CZ gate on the two logical qubits of the $\llbr 4,2,2 \rrbr$ code.}
\vspace*{-0.35cm}
\end{center}
\end{figure}

Although depth or the number of two-qubit gates might appear to be natural metrics for optimization, near-term quantum computers can also benefit from more nuanced metrics depending upon the physical system.
For example, it is now established that the noise in the \emph{IBM Q Experience} computers varies widely among qubits and also with time, and that circuit optimizations might have to be done in regular time intervals in order to exploit the current noise characteristics of the hardware~\cite{Murali-arxiv19}.
In such a scenario, if we need to implement a specific logical operator at the current time, and if it is the case that some specific qubits or qubit-links in the system are particularly unreliable, then it might be better to sacrifice depth and identify an equivalent logical operator that avoids those qubits or qubit-links (if possible).
As an example, for the well-known $\llbr 4,2,2 \rrbr$ code~\cite{Gottesman-phd97,Chao-arxiv17b}, whose stabilizer group is generated as $S = \langle X_1 X_2 X_3 X_4, Z_1 Z_2 Z_3 Z_4 \rangle$, two implementations of the logical controlled-$Z$ ($\lcz{1}{2}$) operation on the two logical qubits are shown in Fig.~\ref{fig:cz_412}.
The logical Pauli operators in this case are $\lX_1 = X_1 X_2, \lX_2 = X_1 X_3, \lZ_1 = Z_2 Z_4, \lZ_2 = Z_3 Z_4$.

Assuming that single-qubit gates do not contribute to complexity (or difficulty of implementation), we observe that both choices have the same number of two-qubit gates and depth.
More interestingly, we see that the second choice completely avoids the first physical qubit while realizing the same logical CZ operation.
Therefore, if either the first qubit itself has poor fidelity or coupling to it does, then clearly the second choice is more appropriate.
Preliminary experiments on the IBM system confirm this advantage when qubits are mapped appropriately.
Note that even if we use a QECC that protects a single qubit but has a transversal CZ implementation, i.e., the logical CZ is a CZ between corresponding physical qubits in two \emph{separate} code blocks, this incurs a larger overhead than the above scheme. 
We identified this example by using our open-source implementation of our LCS algorithm, that is available at: \url{https://github.com/nrenga/symplectic-arxiv18a}.
In order to identify (or construct) more interesting codes that exhibit a ``rich'' set of choices for each logical operator, one needs a better understanding of the geometry of the space of symplectic solutions. 
We believe this is an important open problem arising from our work.



For near-term \emph{NISQ} (Noisy Intermediate-Scale Quantum~\cite{Preskill-nisq18}) era of quantum computers, a lot of current research is focused on equipping compilers with routines that optimize circuits for depth and two-qubit gates, and the mapping of qubits from the algorithm to the hardware, while all taking into account the specific characteristics and noise in the hardware~\cite{Paler-arxiv18,Shi-arch19,Murali-micro19,Nash-arxiv19,Murali-arxiv19}.
Although employing QECCs is considered to be beyond the NISQ regime, exploiting simple codes such as the $\llbr 4,2,2 \rrbr$ code and using post-selection provides increased reliability than uncoded computation (as Harper and Flammia have demonstrated~\cite{Harper-prl19,Linke-sciadv17}). 
Therefore, our efficient LCS algorithm might find an application in such quantum compilers, where the utility is to determine the best physical realization of a logical operator with respect to current system characteristics.
Specifically, this allows \emph{dynamic compilation} (i.e., during program execution) that could provide significant reliability gains in practice.

In light of such applications, our software currently allows one to determine only one physical realization in cases where the number of solutions is prohibitively large, specifically for QECCs with large-dimension stabilizers $(r = m-k \gg 1)$. 
However, this single solution does not come with any explicit guarantees regarding depth or number of two-qubit gates or avoiding certain physical qubits.
Therefore, even developing heuristics to directly optimize for a ``good enough'' solution, instead of assembling all solutions and searching over them, will have a significant impact on the efficiency of compilers.

\subsection{QECCs for Universal Quantum Computation}

Physical single-qubit rotation gates on trapped-ion qubits are natural, reliable and have a long history~\cite{Ozeri-cphy11}.
Recently, it has also been observed that small-angle M{\o}lmer-S{\o}rensen gates, i.e., $XX_{ij}(\theta) = \cos\frac{\theta}{2} \cdot I_4 - \imath \sin\frac{\theta}{2} \cdot X_i X_j$ for small $\theta$, are more reliable than the maximally-entangling $XX_{ij}(\frac{\pi}{2})$ gate~\cite{Nam-arxiv19}.
Since these are the primitive operations in trapped-ion systems~\cite{Linke-nas17}, codes that support a transversal $T = \text{diag}(1,\exp(\frac{\imath\pi}{4}))$ gate, such as the tri-orthogonal codes mentioned earlier, could be directly used for computation rather than being dedicated for expensive magic state distillation~\cite{Bravyi-pra12,Haah-quantum17,Haah-quantum17b,Gidney-arxiv18}.
However, it is well-known that there exists no single QECC that supports a universal set of gates where all of them have a transversal implementation at the logical level~\cite{Zeng-it07,Eastin-prl09,Newman-arxiv17}.
Therefore, there is a natural tradeoff between exploiting transversality for logical non-Clifford operations versus Clifford operations.

Indeed, this will be a realistic alternative only if the logical Clifford operations on these codes are ``error-resilient'', by which we mean that for at least constant-depth circuits, the most likely errors remain correctable and do not propagate catastrophically through the Clifford sections of these logical circuits.
For this purpose, our LCS algorithm can be a supportive tool to investigate properties of stabilizer QECCs that guarantee error-resilience of their logical Clifford operators.
Note that constant-depth circuits have been shown to provide a quantum advantage over classical computation~\cite{Bravyi-science18}.
In fact, it has been shown that the advantage persists even if those circuits are noisy~\cite{Bravyi-arxiv19}, and the proof involves a QECC which admits constant-depth logical Cliffords.


\subsection{Organization}

The paper is organized as follows.
Section~\ref{sec:operators} discusses the connection between quantum computation and the binary symplectic group, which forms the foundation for this work.
Section~\ref{sec:logical_clifford_synthesis} begins by outlining the process of finding logical Clifford gates through a demonstration for the $\llbr 6,4,2 \rrbr$ CSS code\cite{Gottesman-arxiv97,Chao-arxiv17b}. 
Then the general case of stabilizer codes is discussed rigorously via four theorems and our LCS algorithm.
Finally, Section~\ref{sec:conclusion} concludes the paper.
%
Appendix~\ref{sec:trung_proof} discusses the proof of Theorem~\ref{thm:Trung}, and 
Appendix~\ref{sec:alg2_matlab} provides the source code for Algorithm~\ref{alg:symp_lineq_all} with extensive comments.

\section{Physical and Logical Operators}
\label{sec:operators}

Quantum error-correcting codes (QECCs) protect qubits involved in quantum computation.
In this section, we summarize the mathematical framework introduced in~\cite{Calderbank-physreva96, Calderbank-it98*2, Gottesman-phd97, Gottesman-arxiv97} and described in more detail in~\cite{Gottesman-arxiv09,Rengaswamy-arxiv18*2}. 
%
Mathematically, an $m$-qubit system is treated as a Hilbert space with dimension $N=2^m$.
Universal quantum computation requires the ability to implement (within a specified tolerance) quantum operations represented by the group of $N \times N$ unitary matrices acting on this space.
In this paper, we are primarily concerned with the unitary operators in the Clifford group.

\emph{Notation:}
Let $\mathbb{R}$ denote the field of real numbers, $\mathbb{C}$ denote the field of complex numbers, and $\mathbb{F}_2$ denote the binary field.
We will consider vectors over $\mathbb{F}_2$ to be row vectors and vectors over $\mathbb{R}$ or $\mathbb{C}$ to be column vectors.
Vectors over $\mathbb{C}^N$ with $N=2^m$ will be indexed by elements of $\mathbb{F}_2^m$ in the natural binary order $00\cdots 00, 00\cdots 01, \ldots, 11\cdots 11$ (rather than $\{1,2,\ldots,N\}$).
Thus, for $v \in \mathbb{F}_2^m$, let $e_v \in \mathbb{C}^N$ denote the standard basis vector associated with index $v$, i.e., $e_v = \ket{v}$ is all-zero except for a $1$ in the entry indexed by $v$.

\subsection{Pauli Matrices and the Symplectic Inner Product}
\label{sec:heisenberg_weyl}

For a single qubit, we have $m=1$ and a quantum pure state is a vector in the $N=2$ dimensional Hilbert space $\mathbb{C}^2$.
A pure quantum state $\hat{u} \in \mathbb{C}^2$ is a unit-length superposition of the two states ${e_0 \triangleq [1,0]^T}, {e_1 \triangleq [0,1]^T}$ that form the \emph{computational basis}.
Thus, $\hat{u} = \alpha e_0 + \beta e_1$, where $\alpha, \beta \in \mathbb{C}$ satisfy $|\alpha|^2 + |\beta^2| = 1$.
The \emph{Pauli matrices} for a single qubit system are the $2 \times 2$ identity matrix $I_2$,
\begin{align}
X \triangleq 
\begin{bmatrix}
0 & 1 \\
1 & 0
\end{bmatrix} , \ 
Z \triangleq 
\begin{bmatrix}
1 & 0 \\
0 & -1
\end{bmatrix} , \  
Y \triangleq \imath XZ = 
\begin{bmatrix}
0 & -\imath \\
\imath & 0
\end{bmatrix}.
\end{align} 
We note that the Pauli matrices form a basis over $\mathbb{C}$ for all $2\times 2$ complex matrices.
Thus, any single qubit unitary operator (such as an error) can be written as a linear combination of Pauli matrices.
One can also express any pure quantum state as
$\hat{u} = ( \alpha_0 I_2 + \alpha_1 X + \imath \alpha_2 Z + \alpha_3 Y )\ e_0, \ \text{where}\ \alpha_i \in \mathbb{R}$.

For an $m$-qubit system, we work in the $N=2^m$ dimensional Hilbert space $\mathbb{C}^N$ and a pure quantum state $\hat{u}$ is a unit-length vector in this space.
The computational basis vectors $\{ e_v \in \mathbb{C}^N | (v_1,\ldots,v_m) \in\mathbb{F}_2^m \}$ are defined by the Kronecker product $e_v \triangleq e_{v_1} \otimes e_{v_2} \otimes \cdots \otimes e_{v_m}$.
Thus a pure quantum state can be written as
$\hat{u} = \sum_{v\in \mathbb{F}_2^m} \alpha_v e_v$, where $\sum_{v\in \mathbb{F}_2^m} |\alpha_v|^2 = 1$. 

Let $N \triangleq 2^m$. 
Given row vectors $a,b \in \mathbb{F}_2^m$, define the $m$-fold Kronecker product 
\begin{align}
\label{eq:d_ab}
D(a,b) \triangleq X^{a_1} Z^{b_1} \otimes \cdots \otimes X^{a_m} Z^{b_m} \in \mathbb{U}_N,
\end{align}
where $\mathbb{U}_N$ denotes the group of all $N \times N$ unitary operators.
The \emph{Heisenberg-Weyl group} $HW_N$ (also called the \emph{$m$-qubit Pauli group}) consists of all operators $\imath^{\kappa} D(a,b)$, where $\kappa \in \mathbb{Z}_4 \triangleq \{ 0,1,2,3 \} $.
The order is $|HW_N| = 4N^2$ and the \emph{center} of this group is $\langle \imath I_N \rangle = \{ I_N, \imath I_N, -I_N, -\imath I_N \}$, where $I_N$ is the $N \times N$ identity matrix.
Since $XZ = -ZX$,
\begin{align}
\label{eq:hw_commute}
D(a,b) D(a',b') & = (-1)^{a' b^T + b' a^T} D(a',b') D(a,b) \\
  & = (-1)^{a'b^T} D(a+a',b+b'),
\end{align}
and $D(a,b)^T = (-1)^{ab^T} D(a,b)$.
The \emph{symplectic inner product} in $\mathbb{F}_2^{2m}$ is defined as 
\begin{align}
\syminn{[a,b]}{[a',b']} \triangleq a' b^T + b' a^T & = [a,b]\ \Omega \ [a',b']^T\ (\bmod\ 2) , \nonumber \\ 
\text{where}\quad \Omega & \triangleq 
\begin{bmatrix}
0 & I_m \\ 
I_m & 0
\end{bmatrix}.
\end{align}
Two operators $D(a,b)$ and $D(a',b')$ commute if and only if $\syminn{[a,b]}{[a',b']} = 0$.
%
The homomorphism $\gamma \colon HW_N \rightarrow \mathbb{F}_2^{2m}$ defined by
$\gamma(\imath^{\kappa} D(a,b)) \triangleq [a,b] \ \text{for\ all} \ \kappa \in \mathbb{Z}_4$
has kernel $\langle \imath I_N \rangle$, which allows us to represent elements of $HW_N$ (up to multiplication by scalars) as binary vectors.
Since $Y=\imath XZ$ is Hermitian but $XZ$ is not, an additional factor of $\imath$ is required to make $D(a,b)$ Hermitian for each $i \in \{1,\ldots,m\}$ where $a_i b_i = 1$.
Hence, the matrix
\begin{align}
E(a,b) \triangleq \imath^{ab^T \bmod 4} D(a,b)
\end{align}
is Hermitian and $E(a,b)^2 = I_N$, because $X^2 = Z^2 = Y^2 = I_2$.
Given $[a,b], [a',b'] \in \mathbb{F}_2^{2m}$, it can be shown that
\begin{align}
\label{eq:Eab}
E(a,b) E(a',b') & = (-1)^{a' b^T + b' a^T} E(a',b') E(a,b) \\
  & = \imath^{a'b^T - b'a^T} E(a+a',b+b'),
\end{align}
where the exponent and the sums $a+a', b+b'$ are computed modulo $4$ (see~\cite{Rengaswamy-pra19} for the extended definition of $E(a,b)$).

\subsection{Stabilizer Codes}
\label{sec:stabilizer_codes}

We use commutative subgroups of $HW_N$ to define resolutions of the identity.
A \emph{stabilizer} group is a commutative subgroup $S$ of $HW_N$ generated by commuting Hermitian matrices $\pm E(a,b)$, with the additional property that if $E(a,b) \in S$ then $-E(a,b) \notin S$.
Recall that an operator is an orthogonal projection onto its range iff it is idempotent and Hermitian.
Since $E(a,b)^2 = I_N$ for all $a,b\in \mathbb{F}_2^m$, the operator $\frac{I_N \pm E(a,b)}{2}$ is an orthogonal projection onto the $\pm 1$ eigenspace of $E(a,b)$, respectively.
Also, the eigenvalues of each $E(a,b)$ are $\pm 1$ with algebraic multiplicity $N/2$.

Since all elements of $S$ are commuting Hermitian unitary matrices, they can be simultaneously diagonalized with respect to a common orthonormal basis.
We refer to such a basis as the \emph{common eigenbasis} or simply the \emph{eigenbasis} of the subgroup $S$.
In addition, if the subgroup $S$ is generated by $E(a_i,b_i), i = 1,\ldots,r$, then the operator
$\frac{1}{2^r} \prod_{i=1}^{r} (I_N + E(a_i,b_i))$
is an orthogonal projection onto the $2^{k}$-dimensional subspace $V(S)$ fixed pointwise by $S$, i.e., the $+1$ eigenspace of $S$, where $k \triangleq m - r$.
Mathematically, $V(S) \triangleq \{ \ket{\psi} \in \mathbb{C}^N \mid g \ket{\psi} = \ket{\psi}\  \text{for\ all} \ g \in S \}$.
The subspace $V(S)$ is called the $\llbr m,k \rrbr$ \emph{stabilizer code} determined by $S$, where the notation $\llbr m,k \rrbr$ indicates that $V(S)$ encodes $k$ \emph{logical} qubits into $m$ \emph{physical} qubits.
The extended notation $\llbr m,k,d \rrbr$ is used to denote that any undetectable error on the code must act on at least $d$ qubits, i.e., $d$ is the \emph{(minimum) distance} of the stabilizer code.

Let $\gamma(S)$ denote the subspace of $\mathbb{F}_2^{2m}$ formed by the binary representations of the elements of $S$ under the homomorphism $\gamma$.
A generator matrix for $\gamma(S)$ is
$G_S \triangleq [a_i, b_i]_{i = 1,\ldots,r}$ and we have $G_S\, \Omega\, G_S^T = 0_r$,
where $0_r$ is the $r \times r$ all-zero matrix (the subscript is often neglected).
The condition $G_S\, \Omega\, G_S^T = 0$ encodes the fact that elements of $S$ must pairwise commute.

Given a stabilizer $S$ with generators $E(a_i,b_i), i = 1,\ldots,r$, we can define $2^r$ subgroups $S_{\epsilon_1 \cdots \epsilon_r}$ where the index $(\epsilon_1 \cdots \epsilon_r)$ represents that $S_{\epsilon_1 \cdots \epsilon_r}$ is generated by $\epsilon_i E(a_i,b_i)$, $\epsilon_i \in \{ \pm 1 \}$.
Note that
\begin{align}
\label{eq:stab_projectors}
\Pi_{\epsilon_1 \cdots \epsilon_r} \triangleq \frac{1}{2^r} \prod_{i=1}^{r} (I_N + \epsilon_i E(a_i,b_i))
\end{align}
is the orthogonal projector onto $V(S_{\epsilon_1 \cdots \epsilon_r})$ and the sum
$\sum_{(\epsilon_1, \ldots, \epsilon_r) \in \{ \pm 1 \}^r} \Pi_{\epsilon_1 \cdots \epsilon_r} = I_N$
defines a resolution of the identity.
In quantum error correction, it is sufficient to correct Pauli errors (i.e., elements in $HW_N$) because general errors can be approximated by linear combinations of them~\cite{Ekert-physrevlett96}.
Also, the elements of $HW_N$, acting via conjugation, permute the subgroups $S_{\epsilon_1 \cdots \epsilon_r}$.
Given an $\llbr m,k \rrbr$ stabilizer code, it is possible to perform encoded quantum computation in any of the subspaces $V(S_{\epsilon_1 \cdots \epsilon_r})$ by synthesizing appropriate logical operators.
If we think of these subspaces as \emph{threads}, then a computation starts in one thread and jumps to another when an error (from $HW_N$) occurs.
QECCs enable error control by identifying the jump that the computation has made.
Identification makes it possible to adjust future operations in the computation instead of returning to the initial subspace and restarting the computation.
The idea of tracing these threads is called as \emph{Pauli frame tracking}~\cite{Chamberland-quantum17}. 

\subsection{The Clifford Group and Symplectic Matrices}
\label{sec:clifford_gp}

The Clifford group $\text{Cliff}_{N}$ consists of all unitary matrices $g \in \mathbb{C}^{N \times N}$ for which $g D(a,b) g^{\dagger} \in HW_N$ for all $D(a,b) \in HW_N$, where $g^{\dagger}$ is the conjugate transpose of $g$~\cite{Gottesman-arxiv09}. 
$\text{Cliff}_N$ is the \emph{normalizer} of $HW_N$ in the unitary group $\mathbb{U}_N$, so it contains $HW_N$. 
Note that by definition $\text{Cliff}_N$ has an infinite center consisting of $\mathbb{U}(1) \triangleq \{ e^{\imath \theta} I_N; \theta \in \mathbb{R} \}$, but it can be made finite by first taking the quotient group $\text{Cliff}_N/\mathbb{U}(1)$ and then including multiples of just the phase $e^{\imath \pi/4}$, which contributes a factor $8$ in its size~\cite{Calderbank-it98*2}, i.e., $|\text{Cliff}_N| =  8 \cdot 2^{m^2 + 2m} \prod_{j=1}^{m} (4^j - 1)$.
We regard operators in $\text{Cliff}_N$ as \emph{physical} operators acting on quantum states in $\mathbb{C}^N$, to be implemented by quantum circuits.
Every operator $g \in \text{Cliff}_N$ induces an automorphism of $HW_N$ by conjugation.
Note that the inner automorphisms induced by matrices in $HW_N$ preserve every conjugacy class $\{ \pm D(a,b) \}$ and $\{ \pm \imath D(a,b) \}$, because~\eqref{eq:hw_commute} implies that elements in $HW_N$ either commute or anti-commute.
The automorphism induced by an element $g \in \text{Cliff}_{N}$ satisfies (see~\cite{Rengaswamy-arxiv18*2} for a proof)
\begin{align}
\label{eq:symp_action}
g E(a,b) g^{\dagger} = \pm E\left( [a,b] F_g \right), \ \text{where} \ \ F_g = 
\begin{bmatrix}
A_g & B_g \\
C_g & D_g
\end{bmatrix}. 
\end{align}
Since conjugation by $g$ respects commutativity in $HW_N$, the matrix $F_g$ preserves symplectic inner products: 
$\syminn{[a,b]F_g}{[a',b']F_g}$ $= \syminn{[a,b]}{[a',b']}$. 
This implies that $F_g$ satisfies $F_g \Omega F_g^T = \Omega$.
We say that $F_g$ is a \emph{binary symplectic matrix}, and express the symplectic property $F_g \Omega F_g^T = \Omega$ as
$A_g B_g^T = B_g A_g^T, \ C_g D_g^T = D_g C_g^T, \ A_g D_g^T + B_g C_g^T = I_m$.
Let $\text{Sp}(2m,\mathbb{F}_2)$ denote the group of symplectic $2m \times 2m$ matrices over $\mathbb{F}_2$.
The homomorphism $\phi \colon \text{Cliff}_N \rightarrow \text{Sp}(2m,\mathbb{F}_2)$ defined by
$\phi(g) \triangleq F_g $
is surjective with kernel $\langle HW_N, \mathbb{U}(1) \rangle$, and every Clifford operator maps down to a symplectic matrix $F_g$.
Thus, $HW_N$ is a normal subgroup of $\text{Cliff}_N$ and $\text{Cliff}_N/\langle HW_N, \mathbb{U}(1) \rangle \cong \text{Sp}(2m,\mathbb{F}_2)$.
This implies that the size is $|\text{Sp}(2m,\mathbb{F}_2)| = 2^{m^2} \prod_{j=1}^{m} (4^j - 1)$ (also see~\cite{Calderbank-it98*2}).
Table~\ref{tab:std_symp} lists elementary symplectic transformations $F_g$, that generate the binary symplectic group $\text{Sp}(2m,\mathbb{F}_2)$, and the corresponding unitary automorphisms $g \in \text{Cliff}_N$, which together with $HW_N$ generate $\text{Cliff}_N$ 
(see~\cite[Appendix I]{Rengaswamy-arxiv18*2}).
Some important circuit identities involving these operators are listed in~\cite{Rengaswamy-arxiv18*2}.

\begin{table*}
\centering

\caption{\label{tab:std_symp}{\normalfont A universal set of logical operators for $\text{Sp}(2m,\mathbb{F}_2)$ and their corresponding physical operators in $\text{Cliff}_N$ (see Appendix~\ref{sec:elem_symp} 
for a detailed discussion and circuits).
The number of $1$s in $Q$ and $R$ directly relates to the number of gates. 
Here $H_{2^t}$ denotes the Walsh-Hadamard matrix of size $2^t$, $U_t = \text{diag}\left( I_t, 0_{m-t} \right)$ and $L_{m-t} = \text{diag}\left( 0_t, I_{m-t} \right)$. }}

{\small
\begin{tabular}{c|c|c}
Logical Operator $F_{g}$ & Physical Operator $g$ & Circuit Element\\
~ & ~ & ~ \\
\hline
~ & ~ & ~ \\
$\Omega = \begin{bmatrix} 0 & I_m \\ I_m & 0 \end{bmatrix}$ & $H_N = H^{\otimes m} = 
{\renewcommand\arraycolsep{1.35pt}
\frac{1}{(\sqrt{2})^{m}}
\left[
\begin{array}{rr}
1 & 1 \\
1 & -1
\end{array}\right]^{\otimes m}}$ & Transversal Hadamard \\
~ & ~ & ~ \\
$A_Q = \begin{bmatrix} Q & 0 \\ 0 & Q^{-T} \end{bmatrix}$ & $a_Q: \ket{v} \mapsto \ket{v Q}, \, a_Q = \sum_{v \in \mathbb{F}_2^m} \ketbra{vQ}{v}$ & \makecell{Controlled-NOT (CNOT)\\ Qubit Permutation} \\
~ & ~ & ~ \\
\makecell{$T_R = \begin{bmatrix} I_m & R \\ 0 & I_m \end{bmatrix};\, R = R^T$} & \makecell{$t_R\, =\, \text{diag}\left( \imath^{v R v^T \bmod 4} \right)\, =\, \sum_{v \in \mathbb{F}_2^m} \imath^{v R v^T} \ketbra{v}$
} & \makecell{Controlled-$Z$ (CZ)\\ Phase ($P$)}\\
~ & ~ & ~ \\
$G_t = \begin{bmatrix} L_{m-t} & U_t \\ U_t & L_{m-t} \end{bmatrix}$ & $g_t = H_{2^t} \otimes I_{2^{m-t}}$ & Partial Hadamards \\
~ & ~ & ~ \\
\hline
\end{tabular}
}

\end{table*}

In~\cite{Can-sen18}, Can has developed an algorithm that factors a $2m \times 2m$ binary symplectic matrix into a product of at most $6$ elementary symplectic matrices of the type shown in Table~\ref{tab:std_symp}.
The target symplectic matrix maps the dual basis $X_N \triangleq \{ E(a,0) \colon a \in \mathbb{F}_2^m \}, Z_N \triangleq \{ E(0,b) \colon b \in \mathbb{F}_2^m \}$ to a dual basis $X_N', Z_N'$. 
Row and column operations by the elementary matrices return $X_N', Z_N'$ to the original pair $X_N, Z_N$.
This decomposition simplifies the translation of symplectic matrices into circuits (see~\cite[Appendix I]{Rengaswamy-arxiv18*2}), and so we use it in our LCS algorithm.
For completeness, we include the theorem here.

\begin{theorem}[{\hspace{1sp}\cite[Theorem 3.2.1]{Can-sen18}}]
\label{thm:Trung}
Any binary symplectic transformation $F$ can be expressed as
\begin{align*}
F = A_{Q_1} \Omega\, T_{R_1} G_k T_{R_2} A_{Q_2} ,
\end{align*}
as per the notation used in Table~\ref{tab:std_symp}, where invertible matrices $Q_1, Q_2$ and symmetric matrices $R_1,R_2$ are chosen appropriately.
\begin{IEEEproof}
The idea is to perform row and column operations on the matrix $F$ via left and right multiplication by elementary symplectic transformations from Table~\ref{tab:std_symp}, and bring the matrix $F$ to the standard form $\Omega\, T_{R_1} \Omega$ (for details see Appendix~\ref{sec:trung_proof}).
\end{IEEEproof}
\end{theorem}

A closely related algorithm was given earlier by Dehaene and De Moor~\cite{Dehaene-physreva03}.
The elementary symplectic matrices appearing in the product can be related to the Bruhat decomposition of the symplectic group (see~\cite{Maslov-it17}).
When the algorithm is run in reverse it produces a random Clifford matrix, which serves as a ``third-order'' approximation to a random unitary matrix since the Clifford group forms a unitary $3$-design~\cite{Webb-arxiv16}.
This is an instance of the subgroup algorithm~\cite{Diaconis-peis87} for generating uniform random variables.
The algorithm has complexity $O(m^3)$ and uses $O(m^2)$ random bits, which is order optimal given the order of the symplectic group $\text{Sp}(2m,\mathbb{F}_2)$ (cf.~\cite{Koenig-jmp14}).
Our algorithm is similar to that developed by Jones et al.~\cite{Jones-acha13} in that it alternates (partial) Hadamard matrices and diagonal matrices; the difference is that the unitary $3$-design property of the Clifford group provides randomness guarantees.
This also finds application in machine learning (see~\cite{Choromanski-arxiv16} and references therein).

\section{Synthesis of Logical Clifford Operators for Stabilizer Codes}
\label{sec:logical_clifford_synthesis}

Quantum computation in the protected space of an $\llbr m,k \rrbr$ quantum error-correcting code (QECC) requires the translation of logical operators on the $k$ encoded qubits into physical operators on the $m$ code qubits.
In this section, for an $\llbr m,k \rrbr$ stabilizer code, we develop an algorithm that synthesizes \emph{all} physical Clifford realizations of a logical Clifford operator, up to equivalence classes defined by their action on input Pauli operators (which is encoded in their symplectic matrix representation, by~\eqref{eq:symp_action}).
This algorithm makes it possible to optimize the choice of circuit with respect to a metric that is a function of the quantum hardware.
We now outline the algorithm and illustrate the steps using an example where we synthesize a logical controlled-$Z$ gate on the first two logical qubits of the $\llbr 6,4,2 \rrbr$ code~\cite{Gottesman-phd97,Chao-arxiv17b}. 
See~\cite{Rengaswamy-isit18} for discussions on other operators for this code.

\vspace*{1.5mm}

\noindent \emph{Input:} Target Clifford circuit $g$ on the $k$ logical qubits, stabilizers, and logical Paulis.

\noindent \emph{Output:} All Clifford circuits $\bar{g}$ on the $m$ physical qubits that preserve the code space and implement $g$ on the $k$ logical qubits. 

\vspace*{1.5mm}

\noindent \emph{Step 1:} Translate the input into linear constraints on the symplectic matrix $F_{\bar{g}}$ representing $\bar{g}$.

\vspace*{1mm}

The stabilizer group of the $\llbr 6,4,2 \rrbr$ CSS code is $S = \langle X^{\otimes 6}, Z^{\otimes 6} \rangle = \langle E(\vecnot{1},\vecnot{0}), E(\vecnot{0},\vecnot{1}) \rangle$, where $\vecnot{1} = 111111, \vecnot{0} = 000000$.
The logical Pauli operators can be calculated directly~\cite[Section V]{Rengaswamy-arxiv18*2}, or using algorithms developed by Gottesman~\cite{Gottesman-phd97} or Wilde~\cite{Wilde-physreva09}.
These operators are given by $\lX_j = X_1 X_{j+1} = E(\vecnot{e}_1 + \vecnot{e}_{j+1}, \vecnot{0}), \lZ_j = Z_{j+1} Z_6 = E(\vecnot{0}, \vecnot{e}_{j+1} + \vecnot{e}_6), j = 1,2,3,4$, where $\vecnot{e}_j$ is the $j$-th standard basis vector in $\mathbb{F}_2^6$.
We now find a $6$-qubit circuit $\lcz{1}{2}$ on the physical (code) qubits that (i) realizes the $\text{CZ}_{12}$ gate on the logical qubits, i.e., $\lcz{1}{2}$ acts on $\lX_j, \lZ_j$ analogous to how $\text{CZ}_{12}$ acts on $X_j, Z_j$ as stated explicitly in~\eqref{eq:cz}, and (ii) preserves the code space.
Note that satisfying the mathematical condition in (i) does not already guarantee (ii).
The first condition is written as the constraints
\begin{align}
\label{eq:cz}
\lcz{1}{2} \lX_j \lcz{1}{2}^{\dagger} & = 
\begin{cases}
\lX_1 \lZ_2 & \text{if}\ j=1, \\
\lZ_1 \lX_2 & \text{if}\ j=2,\\
\lX_j & \text{if}\ j\neq 1,2,
\end{cases} \nonumber \\
\lcz{1}{2} \lZ_j \lcz{1}{2}^{\dagger} & = \lZ_j \ \text{for\ all} \ j=1,2,3,4 .
\end{align}
The symplectic representation of Clifford elements in~\eqref{eq:symp_action} transforms these conditions into the following linear constraints on the desired symplectic matrix $F_{\lcz{1}{2}}$:
\begin{align}
\label{eq:css642_cz12_a}
[\vecnot{e}_1 + \vecnot{e}_2, \vecnot{0}] F_{\lcz{1}{2}} & = [\vecnot{e}_1 + \vecnot{e}_2, \vecnot{e}_3 + \vecnot{e}_6] , \nonumber \\
[\vecnot{e}_1 + \vecnot{e}_3, \vecnot{0}] F_{\lcz{1}{2}} & = [\vecnot{e}_1 + \vecnot{e}_3, \vecnot{e}_2 + \vecnot{e}_6] , \nonumber \\
[\vecnot{e}_1 + \vecnot{e}_{j+1}, \vecnot{0}] F_{\lcz{1}{2}} & = [\vecnot{e}_1 + \vecnot{e}_{j+1}, \vecnot{0}],\ j = 3,4 , \nonumber \\
[\vecnot{0}, \vecnot{e}_2 + \vecnot{e}_6] F_{\lcz{1}{2}} & = [\vecnot{0}, \vecnot{e}_2 + \vecnot{e}_6], \nonumber \\
[\vecnot{0}, \vecnot{e}_3 + \vecnot{e}_6] F_{\lcz{1}{2}} & = [\vecnot{0}, \vecnot{e}_3 + \vecnot{e}_6], \nonumber \\
[\vecnot{0}, \vecnot{e}_{j+1} + \vecnot{e}_6] F_{\lcz{1}{2}} & = [\vecnot{0}, \vecnot{e}_{j+1} + \vecnot{e}_6],\ j = 3,4.
\end{align}
Constraint (ii) requires that the physical circuit must normalize the stabilizer.
We prove later that any such circuit can be transformed into one that commutes with each stabilizer element, while realizing the same logical operation (see Theorem~\ref{thm:normalize_centralize}).
Note that the requirement is only to preserve the code space, i.e., $\lcz{1}{2}$ must commute with the code projector, but this is equivalent to normalizing the stabilizer since we restrict $\lcz{1}{2}$ to be a physical Clifford operator.
For non-Clifford physical operators, the general approach would be similar to that considered for $Z$-rotations in~\cite{Rengaswamy-arxiv19c}.
Requiring that the circuit centralize the stabilizer yields the constraints
\begin{align}
\label{eq:css642_cz12_b}
[111111,000000] F_{\lcz{1}{2}} & = [111111,000000], \nonumber \\ [000000,111111] F_{\lcz{1}{2}} & = [000000,111111].
\end{align}

\noindent \emph{Step 2:} Find all symplectic solutions.

\vspace*{1mm}

The symplectic constraint $F_{\lcz{1}{2}} \Omega F_{\lcz{1}{2}}^T = \Omega$ is \emph{non-linear}, and in the description of the generic LCS algorithm that follows this example, we show how to use transvections to find all $2^{r(r+1)/2}$ symplectic solutions.
We then translate each solution into a physical Clifford circuit using the decomposition of symplectic matrices as a product of the elementary matrices listed in Table~\ref{tab:std_symp} (see Appendix~\ref{sec:trung_proof} or~\cite{Rengaswamy-arxiv18*2} for details).
For the $\llbr 6,4,2 \rrbr$ code there are $8$ symplectic solutions.
The solution with smallest depth is the elementary symplectic matrix $F_{\lcz{1}{2}} = T_B$, where $B_{23} = B_{32} = B_{26} = B_{62} = B_{36} = B_{63} = 1$ and $B_{ij} = 0$ elsewhere.
The corresponding physical operator $\lcz{1}{2} = \text{diag}\left( \imath^{vBv^T} \right)$ can be decomposed into $\cz{2}{3} \cz{2}{6} \cz{3}{6}$.

\vspace*{1.5mm}

\noindent \emph{Step 3:} Identify any sign violations and find a Pauli matrix to fix the signs while leaving the logical operation undisturbed.

\vspace*{1mm}

The operator $\cz{2}{3} \cz{2}{6} \cz{3}{6}$ commutes with the stabilizer $E(\vecnot{0}, \vecnot{1})$ but not with the stabilizer $E(\vecnot{1}, \vecnot{0})$.
Adding the Pauli operator $Z_6$ fixes the sign and leaves the logical operation undisturbed.
So the final circuit is $\cz{2}{3} \cz{2}{6} \cz{3}{6} Z_6$.

%

\subsection{Symplectic Transvections}
\label{sec:symp_transvec}

\begin{definition}
Given row vector $h \in \mathbb{F}_2^{2m}$, a symplectic transvection is a map $Z_h \colon \mathbb{F}_2^{2m} \rightarrow \mathbb{F}_2^{2m}$ defined by
\begin{align}
\label{eq:symp_transvec}
Z_h(x) \triangleq x + \syminn{x}{h} h = x F_h,\ \text{where}\ F_h \triangleq I_{2m} + \Omega h^T h ,
\end{align}
where $F_h$ is its associated symplectic matrix~\cite{Koenig-jmp14}.
A transvection does not correspond to a single elementary Clifford operator.
\end{definition}

\begin{fact}[{\hspace{1sp}\cite[Theorem 2.10]{Salam-laa08}}]
The symplectic group $\text{Sp}(2m,\mathbb{F}_2)$ is generated by the family of symplectic transvections.
\end{fact}

An important result that is involved in the proof of this fact is the following theorem from~\cite{Salam-laa08,Koenig-jmp14}, which we restate here for $\mathbb{F}_2^{2m}$ since we will build on this result to state and prove Theorem~\ref{thm:symp_lineq}.

\begin{theorem}
\label{thm:symp_transvec}
Let $x,y \in \mathbb{F}_2^{2m}$ be two non-zero vectors. Then $x$ can be mapped to $y$ by a product of at most two symplectic transvections.
\begin{IEEEproof}
There are two cases: $\syminn{x}{y} = 1$ or $0$.
First assume $\syminn{x}{y} = 1$.
Define $h \triangleq x + y$, so
\begin{align*}
x F_h = Z_h(x) & = x + \syminn{x}{x+y} (x+y) \\
  & = x + \left( \syminn{x}{x} + \syminn{x}{y} \right) (x+y) \\
  & = x + (0+1) (x+y) = y .
\end{align*}
Next assume $\syminn{x}{y} = 0$.
Define $h_1 \triangleq w + y, h_2 \triangleq x + w$, where $w \in \mathbb{F}_2^{2m}$ is chosen such that $\syminn{x}{w} = \syminn{y}{w} = 1$.
Then 
\begin{IEEEeqnarray*}{rCl+x*}
x F_{h_1} F_{h_2} & = & Z_{h_2} \left(x + \syminn{x}{w+y} (w+y) \right) & \\
  & = & (x + w+ y) + \syminn{(x+w)+y}{x+w} (x+w) & \\
  & = & y. &  \IEEEQEDhere
\end{IEEEeqnarray*}
\end{IEEEproof}
\end{theorem}

We will use the above result to propose an algorithm (Algorithm~\ref{alg:transvec}) which determines a symplectic matrix $F$ that satisfies $x_i F = y_i,\ i=1,2,\ldots,t \leq 2m$, where $x_i$ are linearly independent and satisfy $\syminn{x_i}{x_j} = \syminn{y_i}{y_j} \ \text{for\ all} \ i,j \in \{1,\ldots,t\}$.


\subsection{Description of the Generic Logical Clifford Synthesis (LCS) Algorithm}
\label{sec:lcs_algorithm}

The synthesis of logical Paulis by Gottesman~\cite{Gottesman-phd97} and by Wilde~\cite{Wilde-physreva09} exploits symplectic groups over the binary field. 
Building on their work we have demonstrated, using the $\llbr 6,4,2 \rrbr$ code as an example, that the binary symplectic group provides a systematic framework for synthesizing physical implementations of any logical operator in the logical Clifford group $\text{Cliff}_{2^k}$ for stabilizer codes. 
In other words, the symplectic group provides a \emph{control plane} where effects of Clifford operators can be analyzed efficiently.
For each logical Clifford operator, one can obtain all symplectic solutions using the algorithm below.

\begin{enumerate}

\item Collect all the linear constraints on $F$, obtained from the conjugation relations of the desired Clifford operator with the stabilizer generators and logical Paulis, to obtain a system of equations $UF = V$.

\item Then vectorize both sides to get $\left( I_{2m} \otimes U \right) \text{vec}(F) = \text{vec}(V)$.

\item Perform Gaussian elimination on the augmented matrix $\left[ \left( I_{2m} \otimes U \right),\ \text{vec}(V) \right]$.
If $\ell$ is the number of non-pivot variables in the row-reduced echelon form, then there are $2^{\ell}$ solutions to the linear system.

\item For each such solution, check if it satisfies $F \Omega F^T = \Omega$.
If it does, then it is a feasible symplectic solution for $\bar{g}$.

\end{enumerate}

Clearly, this algorithm is not very efficient since $\ell$ could be very large.
Specifically, for codes that do not encode many logical qubits this number will be very large as the system $UF = V$ will be very under-constrained.
We now state and prove two theorems that enable us to determine all symplectic solutions for each logical Clifford operator much more efficiently.

\begin{theorem}
\label{thm:symp_lineq}
Let $x_i, y_i \in \mathbb{F}_2^{2m}, i=1,2,\ldots,t \leq 2m$ be a collection of (row) vectors such that $\syminn{x_i}{x_j} = \syminn{y_i}{y_j}$.
Assume that the $x_i$ are linearly independent.
Then a solution $F \in \text{Sp}(2m,\mathbb{F}_2)$ to the system of equations $x_i F = y_i$ can be obtained as the product of a sequence of at most $2t$ symplectic transvections $F_{h} \triangleq I_{2m} + \Omega h^T h$, where $h \in \mathbb{F}_2^{2m}$ is a row vector.
\begin{IEEEproof}
We will prove this result by induction.
For $i=1$ we can simply use Theorem~\ref{thm:symp_transvec} to find $F_1 \in \text{Sp}(2m,\mathbb{F}_2)$ as follows. 
If $\syminn{x_1}{y_1} = 1$ then $F_1 \triangleq F_{h_1}$ with $h_1 \triangleq x_1 + y_1$, or if $\syminn{x_1}{y_1} = 0$ then $F_1 \triangleq F_{h_{11}} F_{h_{12}}$ with $h_{11} \triangleq w_1 + y_1, h_{12} \triangleq x_1 + w_1$, where $w_1$ is chosen such that $\syminn{x_1}{w_1} = \syminn{y_1}{w_1} = 1$.
In any case $F_1$ satisfies $x_1 F_1 = y_1$.
Next consider $i = 2$.
Let $\tilde{x}_2 \triangleq x_2 F_1$ so that $\syminn{x_1}{x_2} = \syminn{y_1}{y_2} = \syminn{y_1}{\tilde{x}_2}$, since $F_1$ is symplectic and hence preserves symplectic inner products.

Similar to Theorem~\ref{thm:symp_transvec} we have two cases: $\syminn{\tilde{x}_2}{y_2} = 1$ or $0$.
For the former, we set $h_2 \triangleq \tilde{x}_2 + y_2$ so that we clearly have $\tilde{x}_2 F_{h_2} = Z_{h_2}(\tilde{x}_2) = y_2$ (see Section~\ref{sec:symp_transvec} for the definition of $Z_h(\cdot)$).
We also observe that
\begin{align*}
y_1 F_{h_2} = Z_{h_2}(y_1) & = y_1 + \syminn{y_1}{\tilde{x}_2 + y_2} (\tilde{x}_2 + y_2) \\
  & = y_1 + (\syminn{y_1}{y_2} + \syminn{y_1}{y_2}) (\tilde{x}_2 + y_2) \\
  & = y_1 .
\end{align*}
Hence in this case $F_2 \triangleq F_1 F_{h_2}$ satisfies $x_1 F_2 = y_1, x_2 F_2 = y_2$.
For the case $\syminn{\tilde{x}_2}{y_2} = 0$ we again find a $w_2$ that satisfies $\syminn{\tilde{x}_2}{w_2} = \syminn{y_2}{w_2} = 1$ and set $h_{21} \triangleq w_2 + y_2, h_{22} \triangleq \tilde{x}_2 + w_2$.
Then by Theorem~\ref{thm:symp_transvec} we clearly have $\tilde{x}_2 F_{h_{21}} F_{h_{22}} = y_2$.
For $y_1$ we observe that
\begin{align*}
& y_1 F_{h_{21}} F_{h_{22}} \\
  & = Z_{h_{22}} \left( Z_{h_{21}}(y_1) \right) \\
  & = Z_{h_{22}} \left( y_1 + \syminn{y_1}{w_2 + y_2} (w_2 + y_2) \right) \\
  & = y_1 + \syminn{y_1}{w_2 + y_2} (w_2 + y_2) + \big( \syminn{y_1}{\tilde{x}_2 + w_2} \\
  & \qquad + \syminn{y_1}{w_2 + y_2} \syminn{w_2 + y_2}{\tilde{x}_2 + w_2} \big) (\tilde{x}_2 + w_2) \\
  & \overset{(a)}{=} y_1 + \syminn{y_1}{w_2 + y_2} (\tilde{x}_2 + y_2) \\
%
  & = y_1 \ \text{if and only if}\ \syminn{y_1}{w_2} = \syminn{y_1}{y_2} ,
\end{align*}
where (a) follows from $\syminn{y_1}{\tilde{x}_2} = \syminn{y_1}{y_2}, \syminn{w_2 + y_2}{\tilde{x}_2 + w_2} = 1+0+0+1 = 0$.
Hence, we pick a $w_2$ such that $\syminn{\tilde{x}_2}{w_2} = \syminn{y_2}{w_2} = 1$ and $\syminn{y_1}{w_2} = \syminn{y_1}{y_2}$, and then set $F_2 \triangleq F_1 F_{h_{21}} F_{h_{22}}$.
Again, for this case $F_2$ satisfies $x_1 F_2 = y_1, x_2 F_2 = y_2$ as well.

By induction, assume $F_{i-1}$ satisfies $x_j F_{i-1} = y_j$ for all $j=1,\ldots,i-1$, where $i \geq 3$.
Using the same idea as for $i=2$ above, let $x_i F_{i-1} = \tilde{x}_i$. 
If $\syminn{\tilde{x}_i}{y_i} = 1$, we simply set $F_i \triangleq F_{i-1} F_{h_i}$, where $h_i \triangleq \tilde{x}_i + y_i$. 
If $\syminn{\tilde{x}_i}{y_i} = 0$, we find a $w_i$ that satisfies $\syminn{\tilde{x}_i}{w_i} = \syminn{y_i}{w_i} = 1$ and $\syminn{y_j}{w_i} = \syminn{y_j}{y_i} \ \forall \ j < i$. 
Then we define $h_{i1} \triangleq w_i + y_i, h_{i2} \triangleq \tilde{x}_i + w_i$ and observe 
\begin{align*}
y_j F_{h_{i1}} F_{h_{i2}} & = Z_{h_{i2}} \left( Z_{h_{i1}}(y_j) \right) \\
  & = y_j + \syminn{y_j}{w_i + y_i} (\tilde{x}_i + y_i) \\
  & = y_j, \ \text{for}\ j < i.
\end{align*}
Again, by Theorem~\ref{thm:symp_transvec}, we clearly have $\tilde{x}_i F_{h_{i1}} F_{h_{i2}} = y_i$.
Hence we set $F_i \triangleq F_{i-1} F_{h_{i1}} F_{h_{i2}}$ in this case. 
In both cases $F_i$ satisfies $x_j F_i = y_j \ \forall \ j=1,\ldots,i$. 
Setting $F \triangleq F_t$ completes the inductive proof and it is clear that $F$ is the product of at most $2t$ symplectic transvections.
\end{IEEEproof}
\end{theorem}


\renewcommand{\algorithmicrequire}{\textbf{Input:}}
\renewcommand{\algorithmicensure}{\textbf{Output:}}

The algorithm defined implicitly by the above proof is stated explicitly in Algorithm~\ref{alg:transvec}.
\begin{algorithm}
\caption{Algorithm to find $F \in \text{Sp}(2m,\mathbb{F}_2)$ satisfying a linear system of equations, using Theorem~\ref{thm:symp_lineq}}	
\label{alg:transvec}
  \begin{algorithmic}[1]	
\REQUIRE $x_i ,y_i \in \mathbb{F}_2^{2m}$ s.t. $\syminn{x_i}{x_j} = \syminn{y_i}{y_j} \ \forall \ i,j \in \{1,\ldots,t\}$. 
\ENSURE $F \in \text{Sp}(2m,\mathbb{F}_2)$ satisfying $x_i F = y_i \ \forall \ i \in \{1,\ldots,t\}$
\IF{$\syminn{x_1}{y_1} = 1$}
	\STATE set $h_1 \triangleq x_1 + y_1$ and $F_1 \triangleq F_{h_1}$.		
\ELSE
	\STATE $h_{11} \triangleq w_1 + y_1, h_{12} \triangleq x_1 + w_1$ and $F_1 \triangleq F_{h_{11}} F_{h_{12}}$.
\ENDIF
		
\FOR{$i = 2,\ldots,t$}
	\STATE Calculate $\tilde{x}_i \triangleq x_i F_{i-1}$ and $\syminn{\tilde{x}_i}{y_i}$.
	\IF{$\tilde{x}_i = y_i$}
		\STATE Set $F_i \triangleq F_{i-1}$. \textbf{Continue}.
	\ENDIF
	\IF{$\syminn{\tilde{x}_i}{y_i} = 1$}
		\STATE Set $h_i \triangleq \tilde{x}_i + y_i, F_i \triangleq F_{i-1} F_{h_i}$.
	\ELSE
		\STATE Find a $w_i$ s.t. $\syminn{\tilde{x}_i}{w_i} = \syminn{y_i}{w_i} = 1$ and $\syminn{y_j}{w_i} = \syminn{y_j}{y_i} \ \forall \ j < i$. 
		\STATE Set $h_{i1} \triangleq w_i + y_i, h_{i2} \triangleq \tilde{x}_i + w_i, F_i \triangleq F_{i-1} F_{h_{i1}} F_{h_{i2}}$.
	\ENDIF
\ENDFOR
\RETURN $F \triangleq F_t$.
\end{algorithmic}
\end{algorithm}

\begin{definition}
\label{def:symp_basis}
A symplectic basis for $\mathbb{F}_2^{2m}$ is a set of pairs $\{ (v_1,w_1), (v_2,w_2), \ldots, (v_m,w_m) \}$ such that $\syminn{v_i}{w_j} = \delta_{ij}$ and $\syminn{v_i}{v_j} = \syminn{w_i}{w_j} = 0$, where $\delta_{ij} = 1$ if $i=j$ and $0$ if $i \neq j$.
\end{definition}

Note that the rows of any matrix in $\text{Sp}(2m,\mathbb{F}_2)$ form a symplectic basis for $\mathbb{F}_2^{2m}$.
There exists a symplectic Gram-Schmidt orthogonalization procedure that can produce a symplectic basis starting from the standard basis for $\mathbb{F}_2^{2m}$ and an additional vector $v \in \mathbb{F}_2^{2m}$ (see~\cite{Koenig-jmp14}).

Now we state our main theorem, which enables one to determine all symplectic solutions for a system of linear equations.

\begin{theorem}
\label{thm:symp_lineq_all}
Let $\{(u_a, v_a),\ a \in \{1,\ldots,m\}\}$ be a collection of pairs of (row) vectors that form a symplectic basis for $\mathbb{F}_2^{2m}$, where $u_a, v_a \in \mathbb{F}_2^{2m}$. 
Consider the system of linear equations $u_i F = u_i', v_j F = v_j'$, where $i \in \mathcal{I} \subseteq \{1,\ldots,m\}, {j \in \mathcal{J} \subseteq \{1,\ldots,m\}}$ and $F \in \text{Sp}(2m,\mathbb{F}_2)$.
Assume that the given vectors satisfy $\syminn{u_{i_1}}{u_{i_2}} = \syminn{u_{i_1}'}{u_{i_2}'} = 0, \syminn{v_{j_1}}{v_{j_2}} = \syminn{v_{j_1}'}{v_{j_2}'} = 0, \syminn{u_{i}}{v_{j}} = \syminn{u_{i}'}{v_{j}'} = \delta_{ij}$, where $i_1,i_2 \in \mathcal{I}, \ j_1,j_2 \in \mathcal{J}$ (since symplectic transformations $F$ must preserve symplectic inner products).
Let $\alpha \triangleq |\bar{\mathcal{I}}| + |\bar{\mathcal{J}}|$, where $\bar{\mathcal{I}}, \bar{\mathcal{J}}$ denote the set complements of $\mathcal{I},\mathcal{J}$ in $\{1,\ldots,m\}$, respectively.
Then there are $2^{\alpha(\alpha+1)/2}$ solutions $F$ to the given linear system, and they can be enumerated systematically.
\begin{IEEEproof}
By the definition of a symplectic basis (Definition~\ref{def:symp_basis}), we have $\syminn{u_a}{v_b} = \delta_{ab}$ and $\syminn{u_a}{u_b} = \syminn{v_a}{v_b} = 0$, where $a,b \in \{1,\ldots,m\}$.
The same definition extends to any (symplectic) subspace of $\mathbb{F}_2^{2m}$.
The linear system under consideration imposes constraints only on $u_i, i \in \mathcal{I}$ and $v_j, j \in \mathcal{J}$.
Let $W$ be the subspace of $\mathbb{F}_2^{2m}$ spanned by the symplectic pairs $(u_c,v_c)$ where $c \in \mathcal{I} \cap \mathcal{J}$ and $W^{\perp}$ be its orthogonal complement under the symplectic inner product, i.e., $W \triangleq \langle \{ (u_c,v_c),\ c \in \mathcal{I} \cap \mathcal{J} \} \rangle$ and $W^{\perp} \triangleq \langle \{ (u_d, v_d),\ d \in \bar{\mathcal{I}} \cup \bar{\mathcal{J}} \} \rangle$, where $\bar{\mathcal{I}}, \bar{\mathcal{J}}$ denote the set complements of $\mathcal{I},\mathcal{J}$ in $\{1,\ldots,m\}$, respectively.

Using the result of Theorem~\ref{thm:symp_lineq}, we first compute one solution $F_0$ for the given system of equations.
In the subspace $W$, $F_0$ maps $(u_c,v_c) \mapsto (u_c',v_c')$ for all $c \in \mathcal{I} \cap \mathcal{J}$ and hence we now have $W = \langle \{ (u_c',v_c'),\ c \in \mathcal{I} \cap \mathcal{J} \} \rangle$ spanned by its new basis pairs $(u_c',v_c')$.
However in $W^{\perp}$, $F_0$ maps $(u_d,v_d) \mapsto (u_d',\tilde{v}_d')$ or $(u_d,v_d) \mapsto (\tilde{u}_d',v_d')$ or $(u_d,v_d) \mapsto (\tilde{u}_d',\tilde{v}_d')$ depending on whether $d \in \mathcal{I} \cap \bar{\mathcal{J}}$ or $d \in \bar{\mathcal{I}} \cap \mathcal{J}$ or $d \in \bar{\mathcal{I}} \cap \bar{\mathcal{J}}$, respectively ($d \notin \mathcal{I} \cap \mathcal{J}$  by definition of $W^{\perp}$).
Note however that the subspace $W^{\perp}$ itself is fixed.
We observe that such $\tilde{u}_d'$ and $\tilde{v}_d'$ are not specified by the given linear system and hence form only a particular choice for the new symplectic basis of $W^{\perp}$. 
These can be mapped to arbitrary choices $\tilde{u}_d$ and $\tilde{v}_d$, while fixing other $u_d'$ and $v_d'$, as long as the new choices still complete a symplectic basis for $W^{\perp}$.
Hence, these form the degrees of freedom for the solution set of the given system of linear equations.
The number of such ``free'' vectors is exactly $|\bar{\mathcal{I}}| + |\bar{\mathcal{J}}| = \alpha$.
This can be verified by observing that the number of basis vectors for $W^{\perp}$ is $2 |\bar{\mathcal{I}} \cup \bar{\mathcal{J}}|$ and making the following calculation.
\begin{align*}
& \text{Number\ of\ constrained\ vectors\ in\ the\ new\ basis\ for}\ W^{\perp} \\
  &= |\mathcal{I} \setminus \mathcal{J}| + |\mathcal{J} \setminus \mathcal{I}| \\
  &= |\mathcal{I}| - |\mathcal{I} \cap \mathcal{J}| + |J| - |\mathcal{I} \cap \mathcal{J}| \\
  &= (m - |\bar{\mathcal{I}}|) + (m - |\bar{\mathcal{J}}|) - 2 (m - |\bar{\mathcal{I}} \cup \bar{\mathcal{J}}|) \\
  &= 2 |\bar{\mathcal{I}} \cup \bar{\mathcal{J}}| - (|\bar{\mathcal{I}}| + |\bar{\mathcal{J}}|) \\
  &= 2 |\bar{\mathcal{I}} \cup \bar{\mathcal{J}}| - \alpha .
\end{align*}

Let  $d, d_1, d_2 \in \bar{\mathcal{I}} \cup \bar{\mathcal{J}}$ be indices of some symplectic basis vectors for $W^{\perp}$.
Then, the constraints on free vectors $\tilde{u}_d$ and $\tilde{v}_d$ are that $\syminn{\tilde{u}_{d_1}}{v_{d_2}'} = \syminn{u_{d_1}'}{\tilde{v}_{d_2}} = \syminn{\tilde{u}_{d_1}}{\tilde{v}_{d_2}} = \delta_{d_1 d_2}$ and all other pairs of vectors in the new basis set for $W^{\perp}$ be orthogonal to each other.
In the $d$-th symplectic pair --- $(\tilde{u}_d,v_d')$ or $(u_d',\tilde{v}_d)$ or $(\tilde{u}_d,\tilde{v}_d)$ --- of its new symplectic basis there is at least one free vector --- $\tilde{u}_d$ or $\tilde{v}_d$ or both, respectively.
For the first of the $\alpha$ free vectors, there are $2 |\bar{\mathcal{I}} \cup \bar{\mathcal{J}}| - \alpha$ symplectic inner product constraints (which are linear constraints) imposed by the $2 |\bar{\mathcal{I}} \cup \bar{\mathcal{J}}| - \alpha$ constrained vectors $u_d',v_d'$. 
Since $W^{\perp}$ has (binary) vector space dimension $2 |\bar{\mathcal{I}} \cup \bar{\mathcal{J}}|$ and each linearly independent constraint decreases the dimension by $1$, this leads to $2^{\alpha}$ possible choices for the first free vector.
For the second free vector, there are $\alpha-1$ degrees of freedom as it has an additional inner product constraint from the first free vector.
This leads to $2^{\alpha-1}$ possible choices for the second free vector, and so on.
Therefore, the given linear system has at least $\prod_{\ell=1}^{\alpha} 2^{\ell} = 2^{\alpha(\alpha+1)/2}$ symplectic solutions.

We will now argue that there cannot be more solutions.
The given system of equations can be represented compactly as $UF = V$, where $U, V \in \mathbb{F}_2^{(2m - \alpha) \times 2m}$ and $F$ is symplectic.
Observe that for each valid choice of $V$ the set of symplectic solutions is disjoint, and hence they form a partition of the binary symplectic group $\text{Sp}(2m,\mathbb{F}_2)$.
Therefore, it is enough to show that the product of the number of such valid matrices $V$ and $2^{\alpha(\alpha + 1)/2}$ is equal to the size of $\text{Sp}(2m,\mathbb{F}_2)$.
By defining $k = m - \alpha$, the number of such valid matrices $V$ is given by
\begin{align*}
& \bigg[ (2^{2m} - 1) \cdot (2^{2m-1} - 2^1) \cdot (2^{2m-2} - 2^2) \cdots \\
  & \hspace{5cm} (2^{2m-(m-1)} - 2^{m-1}) \bigg] \\
  & \qquad \times \bigg[ 2^{2m-m} \cdot 2^{2m-(m+1)} \cdots 2^{2m - (m+k-1)} \bigg] \\
  & = \bigg( 2^0 (2^{2m} - 1) \cdot 2^1(2^{2m-2} - 1) \cdot 2^2(2^{2m-4} - 1) \cdots \\
  & \qquad 2^{m-1}(2^{(m+1) - (m-1)} - 1) \cdot 2^m \bigg) \cdot 2^{m-1} \cdots 2^{m-(k-1)} \\
  & = 2^{\frac{m(m+1)}{2} + (k-1)m - \frac{k(k-1)}{2}} \prod_{j=1}^m (4^j - 1).
\end{align*}
The counting in the first line is as follows.
First, we assume without loss of generality that the pairs of rows $i$ and $(m + i)$ of $V$ form a symplectic pair, for $i = 1,\ldots,k$, and the rows $k+1,\ldots,m$ are orthogonal to all rows of $V$ under the symplectic inner product.
More precisely, the inner products between pairs of rows of $V$ must be the same as those between corresponding pairs of rows of $U$.
But, we assume that we can perform a symplectic Gram-Schmidt process on $U$ so that the above assumption is valid.
For the first row of $V$, we can choose any non-zero vector and there are $(2^{2m} - 1)$ of them.
For the second row, we need to restrict to vectors that are orthogonal to the first row, and we need to eliminate the subspace generated by the first row.
Similarly, for the third row until the $m$-th row, we keep restricting to the subspace of vectors orthogonal to all previous rows and eliminate the subspace generated by all previous rows.
For the $(m+1)$-th row, it needs to be orthogonal to all rows starting from the second to the $m$-th, but it needs to have symplectic inner product $1$ with the first row.
Hence the dimension decreases by $m$ from $2m$, but notice that the subspace generated by the first $m$ rows cannot have any vector that has symplectic inner product $1$ with the first row.
Therefore, we need not subtract this subspace and this gives the count $2^{2m - m}$ for the $(m+1)$-th row.
A similar argument can be made for all remaining rows and this completes the argument for counting.
(It is easy to verify that by substituting $k = m$ above we obtain the size of $\text{Sp}(2m,\mathbb{F}_2)$ exactly.)
Now we expand the exponent of $2$ above to obtain $\frac{1}{2}(m^2 + 2mk - k^2 - m + k)$.

Recollect that the size of the symplectic group is $2^{m^2} \prod_{j=1}^m (4^j - 1)$, and we need to check that the number obtained by dividing this by $2^{\alpha(\alpha + 1)/2}$ is equal to the above number, for $\alpha = m - k$.
Since the product $\prod_{j=1}^m (4^j - 1)$ matches with the expression in the count above, we only have to check that the exponents of $2$ match.
Here, the exponent of $2$ is given by 
\begin{align*}
 & m^2 - \frac{(m-k)(m-k+1)}{2} \\
 & = \frac{1}{2} \left( 2m^2 - (m^2 - 2mk + k^2 + m - k) \right) \\
 & = \frac{1}{2} \left(m^2 + 2mk - k^2 -m + k \right),
\end{align*}
which equals the exponent calculated above.
This completes the proof that the given system has exactly $2^{\alpha(\alpha + 1)/2}$ solutions.

Finally, we show how to get each symplectic solution $F$ for the given linear system. 
First form the matrix $A$ whose rows are the new symplectic basis vectors for $\mathbb{F}_2^{2m}$ obtained under the action of $F_0$, i.e., the first $m$ rows are $u_c',u_d',\tilde{u}_d'$ and the last $m$ rows are $v_c',v_d',\tilde{v}_d'$. 
Observe that this matrix is symplectic and invertible.
Then form a matrix $B = A$ and replace the rows corresponding to free vectors with a particular choice of free vectors, chosen to satisfy the conditions mentioned above.
Note that $B$ and $A$ differ in exactly $\alpha$ rows, and that $B$ is also symplectic and invertible.
Determine the symplectic matrix $F' = A^{-1} B$ which fixes all new basis vectors obtained for $W$ and $W^{\perp}$ under $F_0$ except the free vectors in the basis for $W^{\perp}$.
Then this yields a new solution $F = F_0 F'$ for the given system of linear equations.
Note that if $\tilde{u}_d = \tilde{u}_d'$ and $\tilde{v}_d = \tilde{v}_d'$ for all free vectors, where $\tilde{u}_d', \tilde{v}_d'$ were obtained under the action of $F_0$ on $W^{\perp}$, then $F' = I_{2m}$.
Repeating this process for all $2^{\alpha(\alpha+1)/2}$ choices of free vectors enumerates all the solutions for the linear system under consideration.
\end{IEEEproof}
\end{theorem}

\begin{remark}
For any system of symplectic linear equations $x_i F = y_i,\ i=1,\ldots,t$ where the $x_i$ do not form a symplectic basis for $\mathbb{F}_2^{2m}$, we first calculate a symplectic basis $(u_j,v_j), \ j=1,\ldots,m$ using the symplectic Gram-Schmidt orthogonalization procedure discussed in~\cite{Koenig-jmp14}.
Then we transform the given system into an equivalent system of constraints on these basis vectors $u_j,v_j$ and apply Theorem~\ref{thm:symp_lineq_all} to obtain all symplectic solutions.
\end{remark}

The algorithm defined implicitly by the above proof is stated explicitly in Algorithm~\ref{alg:symp_lineq_all}. 
\begin{algorithm}[t]
\caption{Algorithm to determine all $F \in \text{Sp}(2m,\mathbb{F}_2)$ satisfying a linear system of equations, using Theorem~\ref{thm:symp_lineq_all}}
\label{alg:symp_lineq_all}
\begin{algorithmic}[1]

\REQUIRE $u_a,v_b \in \mathbb{F}_2^{2m}$ s.t. $\syminn{u_a}{v_b} = \delta_{ab}$ and $\syminn{u_a}{u_b} = \syminn{v_a}{v_b} = 0$, where $a,b \in \{1,\ldots,m\}$.

$u_i', v_j' \in \mathbb{F}_2^{2m}$ s.t. $\syminn{u_{i_1}'}{u_{i_2}'} = 0, \syminn{v_{j_1}'}{v_{j_2}'} = 0, \syminn{u_{i}'}{v_{j}'} = \delta_{ij}$, where $i,i_1,i_2 \in \mathcal{I}, \ j,j_1,j_2 \in \mathcal{J},\ \mathcal{I}, \mathcal{J} \subseteq \{1,\ldots,m\}$.

\ENSURE $\mathcal{F} \subset \text{Sp}(2m,\mathbb{F}_2)$ such that each $F \in \mathcal{F}$ satisfies $u_i F = u_i' \ \forall\ i \in \mathcal{I}$, and $v_j F = v_j' \ \forall \ j \in \mathcal{J}$.

\STATE Determine a particular symplectic solution $F_0$ for the linear system using Algorithm~\ref{alg:transvec}.

\STATE Form the matrix $A$ whose $a$-th row is $u_a F_0$ and $(m+b)$-th row is $v_b F_0$, where $a,b \in \{1,\ldots,m\}$. 

\STATE Compute the inverse of this matrix, $A^{-1}$, in $\mathbb{F}_2$.

\STATE Set $\mathcal{F} = \phi$ and $\alpha \triangleq |\bar{\mathcal{I}}| + |\bar{\mathcal{J}}|$, where $\bar{\mathcal{I}}, \bar{\mathcal{J}}$ denote the set complements of $\mathcal{I},\mathcal{J}$ in $\{1,\ldots,m\}$, respectively.

\FOR{$\ell = 1,\ldots,2^{\alpha(\alpha+1)/2}$}
	
	\STATE Form a matrix $B_{\ell} = A$.

	\STATE For $i \notin \mathcal{I}$ and $j \notin \mathcal{J}$ replace the $i$-th and $(m+j)$-th rows of $B_{\ell}$ with arbitrary vectors such that $B_{\ell} \Omega B_{\ell}^T = \Omega$ and $B_{\ell} \neq B_{\ell'}$ for $1 \leq \ell' < \ell$. \hfill \text{$\boldsymbol{/\ast}$ See proof of Theorem~\ref{thm:symp_lineq_all} for details or Appendix~\ref{sec:alg2_matlab}} \text{for example \texttt{MATLAB\textsuperscript{\textregistered}} code $\boldsymbol{\ast /}$}
	
	\STATE Compute $F' = A^{-1} B$.
	
	\STATE Add $F_{\ell} \triangleq F_0 F'$ to $\mathcal{F}$.
	
\ENDFOR

\RETURN $\mathcal{F}$

\end{algorithmic}
\end{algorithm}
For a given system of linear (independent) equations, if $\alpha = 0$ then the symplectic matrix $F$ is fully constrained and there is a unique solution.
Otherwise, the system is partially constrained and we refer to a solution $F$ as a \emph{partial} symplectic matrix.

\emph{Example}:
As an application of this theorem, we discuss the procedure to determine all symplectic solutions for the logical controlled-$Z$ gate $\lcz{1}{2}$ discussed at the beginning of this section.
First we define a symplectic basis for $\mathbb{F}_2^{12}$ using the binary vector representation of the logical Pauli operators and stabilizer generators of the $\llbr 6,4,2 \rrbr$ code.
\begin{align}
u_1 \triangleq [110000,000000] \quad &, \quad v_1 \triangleq [000000,010001] , \nonumber \\ 
u_2 \triangleq [101000,000000] \quad &, \quad v_2 \triangleq [000000,001001] , \nonumber \\ 
u_3 \triangleq [100100,000000] \quad &, \quad v_3 \triangleq [000000,000101] , \nonumber \\ 
u_4 \triangleq [100010,000000] \quad &, \quad v_4 \triangleq [000000,000011] , \nonumber \\ 
u_5 \triangleq [111111,000000] \quad &, \quad v_5 \triangleq [000000,000001] , \nonumber \\ 
u_6 \triangleq [100000,000000] \quad &, \quad v_6 \triangleq [000000,111111] .
\end{align}
Note that $v_5$ and $u_6$ do not correspond to either a logical Pauli operator or a stabilizer element but were added to complete a symplectic basis.
Hence we have $\mathcal{I} = \{1,2,3,4,5\}, \mathcal{J} = \{1,2,3,4,6\}$ and $\alpha = 1 + 1 = 2$.
As discussed earlier, we impose constraints on all $u_i,v_j$ except for $i=6$ and $j=5$.
Therefore, as per the notation in the above proof, we have $W \triangleq \langle \{(u_1,v_1), \ldots, (u_4,v_4)\} \rangle$ and $W^{\perp} \triangleq \langle \{(u_5,v_5), (u_6,v_6)\} \rangle$.
Using Algorithm~\ref{alg:transvec} we obtain a particular solution $F_0 = T_B$ where $B$ is given in the beginning of Section~\ref{sec:logical_clifford_synthesis}.
Then we compute the action of $F_0$ on the bases for $W$ and $W^{\perp}$ to get
\begin{align}
u_i F_0 \triangleq u_i',\ v_j F_0 \triangleq v_j',\ i \in \mathcal{I},\ j \in \mathcal{J}, \ \ \text{and} \nonumber \\ 
u_6 F_0 = [100000,000000] \triangleq \tilde{u}_6', \nonumber \\
v_5 F_0 = [000000,000001] \triangleq \tilde{v}_5' ,
\end{align}
where $u_i', v_j'$ are the vectors obtained in~\eqref{eq:css642_cz12_a},~\eqref{eq:css642_cz12_b}.
Then we identify $\tilde{v}_5$ and $\tilde{u}_6$ to be the free vectors and one particular solution is $\tilde{v}_5 = \tilde{v}_5', \tilde{u}_6 = \tilde{u}_6'$.
In this case we have $2^{\alpha} = 2^2 = 4$ choices to pick $\tilde{v}_5$, since we need $\syminn{u_5'}{\tilde{v}_5} = 1$, $\syminn{u_i'}{\tilde{v}_5} = 0$ for $i = 1,2,3,4$, and $\syminn{v_j}{\tilde{v}_5} = 0$ for $j = 1,2,3,4,6$. 
For each such choice of $\tilde{v}_5$, we have $2^{\alpha-1} = 2$ choices for $\tilde{u}_6$.
Next we form the matrix $A$ whose $i$-th row is $u_i'$ and $(6+j)$-th row is $v_j'$, where $i \in \mathcal{I},j \in \mathcal{J}$.
We set the $6$th row to be $\tilde{u}_6'$ and the $11$th row to be $\tilde{v}_5'$.
Then we form a matrix $B = A$ and replace rows $6$ and $11$ by one of the $8$ possible pair of choices for $\tilde{u}_6$ and $\tilde{v}_5$, respectively.
This yields the matrix $F' = A^{-1} B$ and the symplectic solution $F = F_0 F'$.
Looping through all the $8$ choices we obtain the solutions listed in the appendices of~\cite{Rengaswamy-arxiv18*2}.

\begin{theorem}
\label{thm:stabilizer_solutions}
For an $\llbr m,k \rrbr$ stabilizer code, the number of solutions for each logical Clifford operator is $2^{r(r+1)/2}$, ignoring stabilizer degrees of freedom (Remark~\ref{rem:stab_freedom}), where $r = m-k$.
\begin{IEEEproof}
Let $u_i, v_i \in \mathbb{F}_2^{2m}$ represent the logical Pauli operators $\lX_i, \lZ_i$, for $i=1,\ldots,k$, respectively, i.e., $\gamma(\lX_i) = u_i, \gamma(\lZ_i) = v_i$, where $\gamma$ is the map defined in Section~\ref{sec:heisenberg_weyl}.
Since $\lX_i \lZ_i = -\lZ_i \lX_i$ and $\lX_i \lZ_j = \lZ_j \lX_i$ for all $j \neq i$, it is clear that $\syminn{u_i}{v_j} = \delta_{ij}$ for $i,j \in \{1,\ldots,k\}$ and hence they form a partial symplectic basis for $\mathbb{F}_2^{2m}$.
Let $u_{k+1},\ldots,u_m$ represent the stabilizer generators, i.e., $\gamma(S_j) = u_{k+j}$ where the stabilizer group is $S = \langle S_1,\ldots,S_r \rangle$.
Since by definition $\lX_i, \lZ_i$ commute with all stabilizer elements, it is clear that $\syminn{u_i}{u_j} = \syminn{v_i}{u_j} = 0$ for $i \in \{1,\ldots,k\}, j \in \{k+1,\ldots,m\}$.
To complete the symplectic basis we find vectors $v_{k+1},\ldots,v_m$ s.t. $\syminn{u_i}{v_j} = \delta_{ij} \ \forall \ i,j \in \{1,\ldots,m\}$.
Now we note that for any logical Clifford operator, the conjugation relations with logical Paulis yield $2k$ constraints, on $u_i,v_i$ for $i \in \{1,\ldots,k\}$, and the normalization condition on the stabilizer yields $r$ constraints, on $u_{k+1},\ldots,u_m$.
Hence we have $\bar{\mathcal{I}} = \phi, \bar{\mathcal{J}} = \{k+1,\ldots,m\}$, as per the notation in Theorem~\ref{thm:symp_lineq_all}, and thus $\alpha = |\bar{\mathcal{I}}| + |\bar{\mathcal{J}}| = m - k = r$.
\end{IEEEproof}
\end{theorem}

\vspace{0.1cm}

\begin{corollary}
For any logical $k$-qubit Clifford operation on an $\llbr m,k \rrbr$ stabilizer code, there always exists a physical $m$-qubit Clifford circuit that normalizes the stabilizer and realizes the given operation.
Effectively, this identifies the surjection $\text{Cliff}_N \cap \mathcal{N}_{\mathbb{U}_N}(S) \rightarrow \text{Cliff}_{2^k}$ whose kernel is all the physical Cliffords that normalize the stabilizer but realize only the logical identity (see the proof of Theorem~\ref{thm:normalize_centralize} for a method to identify them).
Here, $\mathcal{N}_{\mathbb{U}_N}(S)$ denotes the normalizer of $S$ in the group $\mathbb{U}_N$ of all $m$-qubit unitary operations.
\end{corollary}

Note that, for each symplectic solution, there are multiple decompositions into elementary forms (from Table~\ref{tab:std_symp}) possible; one possibility is given in Theorem~\ref{thm:Trung}.
Although each decomposition yields a different circuit, all of them will act identically on $X_N$ and $Z_N$ under conjugation (see Section~\ref{sec:clifford_gp} for notation).
Once a logical Clifford operator is defined by its conjugation with the logical Pauli operators, a physical realization of the operator could either normalize the stabilizer or centralize it, i.e., fix each element of the stabilizer group under conjugation.
We note here that any obtained normalizing solution can be converted into a centralizing solution. 
While we do not have a well-motivated application for this result yet, we believe this might be useful in Pauli frame tracking~\cite{Chamberland-quantum17} and adapting future logical operations to the current signs.

\vspace{0.1cm}

\begin{theorem}
\label{thm:normalize_centralize}
For an $\llbr m,k \rrbr$ stabilizer code with stabilizer $S$, each physical realization of a given logical Clifford operator that normalizes $S$ can be converted into a circuit that centralizes $S$ while realizing the same logical operation.
\begin{IEEEproof}
Let the symplectic solution for a specific logical Clifford operator $\bar{g} \in \text{Cliff}_N$ that normalizes the stabilizer $S$ be denoted by $F_n$. 
Define the logical Pauli groups $\lX \triangleq \langle \lX_1,\ldots,\lX_{k} \rangle$ and $\lZ \triangleq \langle \lZ_1,\ldots,\lZ_{k} \rangle$. 
Let $\gamma(\lX)$ and $\gamma(\lZ)$ denote the matrices whose rows are $\gamma(\lX_i)$ and $\gamma(\lZ_i)$, respectively, for $i=1,\ldots,k$, where $\gamma$ is the map defined in Section~\ref{sec:heisenberg_weyl}. 
Similarly, let $\gamma(S)$ denote the matrix whose rows are the images of the stabilizer generators under the map $\gamma$.
Then, by stacking these matrices as in the proof of Theorem~\ref{thm:stabilizer_solutions}, we observe that $F_n$ is a solution of the linear system
\begin{align*}
\begin{bmatrix}
\gamma(\lX) \\
\gamma(S) \\
\gamma(\lZ)
\end{bmatrix} F_n = 
\begin{bmatrix}
\gamma(\lX') \\
\gamma(S') \\
\gamma(\lZ')
\end{bmatrix} ,
\end{align*}
where $\lX', \lZ'$ are defined by the conjugation relations of $\bar{g}$ with the logical Paulis, i.e., $\bar{g} \lX_i \bar{g}^{\dagger} = \lX_i', \bar{g} \lZ_i \bar{g}^{\dagger} = \lZ_i'$, and $S'$ denotes the stabilizer group of the code generated by a different set of generators than that of $S$.
Note, however, that as a group $S' = S$.
The goal is to find a different solution $F_c$ that centralizes the stabilizer, i.e. we replace $\gamma(S')$ with $\gamma(S)$ above.

We first find a matrix $K \in \text{GL}(m-k, \mathbb{F}_2)$ such that $K \gamma(S') = \gamma(S)$, which always exists since generators of $S'$ span $S$ as well.
Then we determine a symplectic solution $H$ for the linear system
\begin{align*}
\begin{bmatrix}
\gamma(\lX) \\
\gamma(S) \\
\gamma(\lZ)
\end{bmatrix} H = 
\begin{bmatrix}
\gamma(\lX) \\
K \gamma(S) \\
\gamma(\lZ)
\end{bmatrix} ,
\end{align*}
so that $H$ satisfies $K \gamma(S) = \gamma(S) H$ while fixing $\gamma(\lX)$ and $\gamma(\lZ)$.
Then since $K$ is invertible we can write
\begin{align*}
\begin{bmatrix}
I_{k} &   &         \\
        & K &         \\
        &   & I_{k}
\end{bmatrix}
\begin{bmatrix}
\gamma(\lX) \\
\gamma(S) \\
\gamma(\lZ)
\end{bmatrix} F_n & = 
\begin{bmatrix}
I_{k} &   &         \\
        & K &         \\
        &   & I_{k}
\end{bmatrix}
\begin{bmatrix}
\gamma(\lX') \\
\gamma(S') \\
\gamma(\lZ')
\end{bmatrix} \\
\Rightarrow 
\begin{bmatrix}
\gamma(\lX) \\
\gamma(S) \\
\gamma(\lZ)
\end{bmatrix} H F_n & = 
\begin{bmatrix}
\gamma(\lX') \\
\gamma(S) \\
\gamma(\lZ')
\end{bmatrix} .
\end{align*}
Hence, $F_c \triangleq H F_n$ is a centralizing solution for $\bar{g}$.
Note that there are $2^{r(r+1)/2}$ solutions for $H$, as per the result of Theorem~\ref{thm:stabilizer_solutions} with the operator being the identity operator on the logical qubits, and these produce all centralizing solutions for $\bar{g}$.
\end{IEEEproof}
\end{theorem}

Although any normalizing solution can be converted into a centralizing solution, the optimal solution with respect to a suitable metric need not always centralize the stabilizer.
However, we can always setup the problem of identifying a symplectic matrix, representing the physical circuit, by constraining it to centralize the stabilizer.
The general procedure to determine all symplectic solutions, and their circuits, for a logical Clifford operator for a stabilizer code is summarized in Algorithm~\ref{alg:log_ops}.
For the $\llbr 6,4,2 \rrbr$ CSS code, we employed Algorithm~\ref{alg:log_ops} to determine the solutions listed in the appendices of~\cite{Rengaswamy-arxiv18*2} for each of the standard generating operators for the Clifford group (see Table~\ref{tab:std_symp}).

\begin{algorithm}
\caption{LCS Algorithm to determine all logical Clifford operators (see Section~\ref{sec:operators} for the homomorphisms $\gamma,\phi$)}
\label{alg:log_ops}
\begin{algorithmic}[1]

\STATE Determine the target logical operator $\bar{g}$ by specifying its action on logical Paulis $\lX_i, \lZ_i$~\cite{Gottesman-arxiv09}: $\bar{g} \lX_i \bar{g}^{\dagger} = \lX_i', \bar{g} \lZ_i \bar{g}^{\dagger} = \lZ_i'$ .

\STATE Transform the above relations into linear equations on $F \in \text{Sp}(2m,\mathbb{F}_2)$ using the map $\gamma$ and the result of~\eqref{eq:symp_action}, i.e., $\gamma(\lX_i) F = \gamma(\lX_i'), \gamma(\lZ_i) F = \gamma(\lZ_i')$. 
Add the conditions for normalizing the stabilizer $S$, i.e., $\gamma(S) F = \gamma(S')$.

\STATE Calculate the feasible symplectic solution set $\mathcal{F}$ using Algorithm~\ref{alg:symp_lineq_all} by mapping $\lX_i, S, \lZ_i$ to $u_i, v_i$ as in Theorem~\ref{thm:stabilizer_solutions}.

\STATE Factor each $F \in \mathcal{F}$ into a product of elementary symplectic transformations listed in Table~\ref{tab:std_symp}, possibly using the algorithm given in~\cite{Can-2017a} (which is restated in Theorem~\ref{thm:Trung} here), and compute the physical Clifford operator $\bar{g}$. 

\STATE Check for conjugation of $\bar{g}$ with the stabilizer generators and for the conditions derived in step 1.
If some signs are incorrect, post-multiply by an element from $HW_N$ as necessary to satisfy all these conditions (apply~\cite[Prop. 10.4]{Nielsen-2010} for $S^{\perp} = \langle S, \lX_i, \lZ_i \rangle$, using $\gamma$). 
Since $HW_N$ is the kernel of the map $\phi$, post-multiplication does not change $F$.

\STATE Express $\bar{g}$ as a sequence of physical Clifford gates corresponding to the elementary symplectic matrices obtained from the factorization in step 4 (see 
Appendix~\ref{sec:elem_symp} for the circuits for these matrices).

\end{algorithmic}
\end{algorithm}

The \texttt{MATLAB\textsuperscript{\textregistered}} programs for all algorithms in this paper are available at \url{https://github.com/nrenga/symplectic-arxiv18a}.
We executed our programs on a laptop running the Windows 10 operating system (64-bit) with an Intel\textsuperscript{\textregistered} Core\textsuperscript{\texttrademark} i7-5500U @ 2.40GHz processor and 8GB RAM.
For the $\llbr 6,4,2 \rrbr$ CSS code, it takes about 0.5 seconds to generate all 8 symplectic solutions and their circuits for one logical Clifford operator.
For the $\llbr 5,1,3 \rrbr$ perfect code, it takes about 20 seconds to generate all 1024 solutions and their circuits.
Note that for step 5 in Algorithm~\ref{alg:log_ops}, we use 1-qubit and 2-qubit unitary matrices (from $\text{Cliff}_{2^2}$) to calculate conjugations for the Pauli operator on each qubit, at each circuit element at each depth, and then combine the results to compute the conjugation of $\bar{g}$ with a stabilizer generator or logical Pauli operator.
Owing to our naive implementation, we observe that most of the time is consumed in computing Kronecker products and not in calculating the symplectic solutions.

\begin{remark}
\label{rem:stab_freedom}
Observe that, in our LCS algorithm, we are not taking into account the degrees of freedom provided by stabilizers.
That is, if the logical operator $\bar{g}$ is required to map $\lX_i \mapsto \lX_i'$, then an equivalent condition is to map $\lX_i \mapsto \lX_i' \cdot \boldsymbol{s}$, where $\boldsymbol{s} \in S$ is any stabilizer element for the given code.
A similar statement is true for $\lZ_i \mapsto \lZ_i'$.
An explicit example for this scenario is the $\lcnot{1}{2}$ for the $\llbr 4,2,2 \rrbr$ code with the logical Paulis defined instead as $\lX_1 = X_1 X_2, \lX_2 = X_2 X_4, \lZ_1 = Z_1 Z_3, \lZ_2 = Z_3 Z_4$.
The operation $\lcnot{1}{2}$ can simply be defined as swapping qubits $2$ and $4$, but this maps $\lZ_2 \mapsto Z_2 Z_3 = \lZ_1 \lZ_2 \cdot \bg^Z$, where $\bg^Z = Z_1 Z_2 Z_3 Z_4$, instead of just $\lZ_2 \mapsto \lZ_1 \lZ_2$ as the above algorithm would typically require.

In principle, the LCS algorithm can be easily modified to consider these possibilities, but this significantly increases the computational complexity of the algorithm.
A better understanding of the structure of logical Clifford operators for a given general stabilizer code, or even heuristics developed to identify which degrees of freedom are worth considering for a given code, would greatly improve the quality of solutions produced by the overall algorithm.
\end{remark}

\section{Conclusion}
\label{sec:conclusion}

In this work we have used the binary symplectic group to propose a systematic algorithm for synthesizing physical (Clifford) implementations of logical Clifford operators for any stabilizer code.
This algorithm provides as solutions all symplectic matrices corresponding to the desired logical operator, each of which is subsequently transformed into a circuit by decomposing it into elementary forms.
This decomposition is not unique, and in future work we will address optimization of the synthesis algorithm with respect to circuit complexity, error-resilience, and also other nuanced metrics discussed in the introduction.
For such optimization to be feasible, one might have to explore opportunities for identifying and exploring the algebraic structure hidden in the algorithm, since combinatorially the matrix inversion involved in Algorithm~\ref{alg:symp_lineq_all} could itself form a bottleneck.





\appendices

\section{Elementary Symplectic Transformations and their Circuits}
\label{sec:elem_symp}

In this section we verify that the physical operators listed in Table~\ref{tab:std_symp} are associated with the corresponding symplectic transformation~\cite{Can-2017a}.
Furthermore, we also provide circuits that realize these physical operators (also see~\cite{Dehaene-physreva03}).

Since each physical operator in Table~\ref{tab:std_symp} is a unitary Clifford operator, it is enough to consider their actions on elements of the Heisenberg-Weyl group $HW_N$, where $N=2^m$.
Let $e_v$ be a standard basis (column) vector in $\mathbb{C}^N$ indexed by the vector $v \in \mathbb{F}_2^m$ such that it has entry $1$ in position $v$ and $0$ elsewhere.
More precisely, if $v = [v_1,v_2,\ldots,v_m]$ then $e_v = e_{v_1} \otimes e_{v_2} \otimes \cdots e_{v_m}$, where $e_0 \triangleq \begin{bmatrix} 1 \\ 0 \end{bmatrix} = \ket{0}, e_1 \triangleq \begin{bmatrix} 0 \\ 1 \end{bmatrix} = \ket{1}$.
Hence, we can simply write $e_v = \ket{v} = \ket{v_1} \otimes \cdots \otimes \ket{v_m}$.

\begin{enumerate}

\item $H_N = H^{\otimes m} \colon$ The single-qubit Hadamard operator $H \triangleq \frac{1}{\sqrt{2}} \begin{bmatrix}
1 & 1 \\
1 & -1
\end{bmatrix}$ satisfies $H X H^{\dagger} = Z, H Z H^{\dagger} = X$. 
Hence, the action of $H_N$ on a $HW_N$ element $D(a,b)$ is given by
\begin{align}
H_N D(a,b) H_N^{\dagger} & = H_N D(a,0) D(0,b) H_N^{\dagger} \\
  & = (H_N D(a,0) H_N^{\dagger}) \nonumber \\
  & \qquad \quad (H_N D(0,b) H_N^{\dagger}) \\
  & = D(0,a) D(b,0) \\
  & = (-1)^{ab^T} D(b,a) \\
\Rightarrow  H_N D(a,b) H_N^{\dagger} & = (-1)^{ab^T} D \left( [a,b] \Omega \right). 
\end{align}
The circuit for $H_N$ is just $H$ applied to each of the $m$ qubits.

\item $GL(m,\mathbb{F}_2) \colon$ Each non-singular $m \times m$ binary matrix $Q$ is associated with a symplectic transformation $A_Q$ given by 
\begin{align}
A_Q = 
\begin{bmatrix} 
Q & 0 \\ 
0 & Q^{-T} 
\end{bmatrix} ,
\end{align}
where $Q^{-T} = (Q^T)^{-1} = (Q^{-1})^T$.
The matrix $Q$ is also associated with the unitary operator $a_Q$ which realizes the mapping $e_v \mapsto e_{vQ}$.
We verify this as follows.
Note that $D(c,0) e_v = e_{v+c}$ and $D(0,d) e_v = (-1)^{vd^T} e_v$.
We calculate $(a_Q D(c,d) a_Q^{\dagger}) e_v$
\begin{align}
& = a_Q D(c,0) D(0,d) e_{vQ^{-1}} \\
                               & = a_Q (-1)^{cd^T} D(0,d) D(c,0) e_{vQ^{-1}} \\
                               & = (-1)^{cd^T} a_Q (-1)^{(vQ^{-1}+c) d^T} e_{vQ^{-1}+c} \\
                               & = (-1)^{cd^T} (-1)^{(v+cQ) Q^{-1}d^T} e_{v+cQ} \\
                               & = (-1)^{cd^T} D(0,d(Q^{-1})^T) D(cQ,0) e_v \\
                               & = D(cQ,dQ^{-T}) e_v \\
                               & = D \left( [c,d] A_Q \right) e_v.
\end{align}
Since the operator $a_Q$ realizes the map $e_v = \ket{v} \mapsto \ket{vQ}$, the circuit for the operator is equivalent to the binary circuit that realizes $v \mapsto vQ$.
Evidently, this elementary transformation encompasses CNOT operations and qubit permutations.
For the latter, $Q$ will be a permutation matrix.
Note that if $a_Q$ preserves the code space of a CSS code then the respective permutation must be in the automorphism group of the constituent classical code.
This is the special case that is discussed in detail by Grassl and Roetteler in~\cite{Grassl-isit13}.

\vspace{0.1cm}
For a general $Q$, one can use the LU decomposition over $\mathbb{F}_2$ to obtain $P_{\pi} Q = LU$, where $P_{\pi}$ is a permutation matrix, $L$ is lower triangular and $U$ is upper triangular.
Note that $L_{ii} = U_{ii} = 1 \ \forall \ i \in \{1,\ldots,m\}$.
Then the circuit for $Q$ first involves the permutation $P_{\pi}^T$ (or $\pi^{-1}$), then CNOTs for $L$ with control qubits in the order $1,2,\ldots,m$ and then CNOTs for $U$ with control qubits in reverse order $m,m-1,\ldots,1$.
The order is important because an entry $L_{ji} = 1$ implies a CNOT gate with qubit $j$ controlling qubit $i$ (with $j > i$), i.e, $\cnot{j}{i}$, and similarly $L_{kj} = 1$ implies the gate $\cnot{k}{j}$ (with $k > j$).
Since the gate $\cnot{j}{i}$ requires the value of qubit $j$ \emph{before} it is altered by $\cnot{k}{j}$, it needs to be implemented first.
A similar reasoning applies to the reverse order of control qubits for $U$.

\item $t_R = \text{diag}\left( \imath^{vRv^T} \right) \colon$ Each symmetric matrix $R \in \mathbb{F}_2^{m \times m}$ is associated with a symplectic transformation $T_R$ given by
\begin{align}
T_R = 
\begin{bmatrix} 
I_m & R \\ 
0 & I_m 
\end{bmatrix},
\end{align}
and with a unitary operator $t_R$ that realizes the map $e_v \mapsto \imath^{vRv^T} e_v$.
We now verify that conjugation by $t_R$ induces $T_R$.
We calculate $(t_R D(a,b) t_R^{\dagger}) e_v$ 
\begin{align}
& = \imath^{-vRv^T} t_R (-1)^{ab^T} D(0,b) D(a,0) e_v \\
                               & = \imath^{-vRv^T} (-1)^{ab^T} t_R (-1)^{(v+a)b^T} e_{v+a} \\
                               & = (-1)^{ab^T} \imath^{-vRv^T} (-1)^{(v+a)b^T} \imath^{(v+a)R(v+a)^T} e_{v+a} \\
                               & = (-1)^{ab^T} \imath^{aRa^T} (-1)^{vRa^T + (v+a)b^T} e_{v+a} \\
                               & = (-1)^{ab^T} \imath^{-aRa^T} (-1)^{(v+a)(b+aR)^T} e_{v+a} \\
                               & = (-1)^{ab^T} \imath^{-aRa^T}  D(0,b+aR) D(a,0) e_v \\
                               & = (-1)^{ab^T} \imath^{-aRa^T} (-1)^{a(b+aR)^T} D(a,b+aR) e_v \\
                               & = \imath^{aRa^T} D \left( [a,b] T_R \right) e_v.
\end{align}
Hence, for $E(a,b) \triangleq \imath^{ab^T} D(a,b)$, we have $t_R E(a,b) t_R^{\dagger} = \imath^{ab^T} \imath^{aRa^T} D(a,b+aR) = E \left( [a,b] T_R \right)$ as required.
We derive the circuit for this unitary operator by observing the action of $T_R$ on the standard basis vectors $[\vecnot{e}_1,\vecnot{0}],\ldots,[\vecnot{e}_m,\vecnot{0}]$, $[\vecnot{0},\vecnot{e}_1],\ldots,[\vecnot{0},\vecnot{e}_m]$ of $\mathbb{F}_2^{2m}$, where $i \in \{1,\ldots,m\}$, which captures the effect of $t_R$ on the (basis) elements $X_1,\ldots,X_m$, $Z_1,\ldots,Z_m$ of $HW_N$, respectively, under conjugation.

\vspace{0.1cm}
Assume as the first special case that $R$ has non-zero entries only in its (main) diagonal.
If $R_{ii} = 1$ then we have $[\vecnot{e}_i,\vecnot{0}] T_R = [\vecnot{e}_i, \vecnot{e}_i]$.
This indicates that $t_R$ maps $X_i \mapsto X_i Z_i \approx Y_i$.
Since we know that the phase gate $P_i$ on the $i$-th qubit performs exactly this map under conjugation, we conclude that the circuit for $t_R$ involves $P_i$.
We proceed similarly for every $i \in \{1,\ldots,m\}$ such that $R_{ii} = 1$.

\vspace{0.1cm}
Now consider the case where $R_{ij} = R_{ji} = 1$ (since $R$ is symmetric).
Then we have 
\begin{align}
{[\vecnot{e}_i,\vecnot{0}] T_R = [\vecnot{e}_i,\vecnot{e}_j], \ [\vecnot{e}_j,\vecnot{0}] T_R = [\vecnot{e}_j,\vecnot{e}_i]}.
\end{align}
This indicates that $t_R$ maps $X_i \mapsto X_i Z_j$ and $X_j \mapsto Z_i X_j$.
Since we know that the controlled-$Z$ gate ${\rm CZ}_{ij}$ on qubits $(i,j)$ performs exactly this map under conjugation, we conclude that the circuit for $t_R$ involves ${\rm CZ}_{ij}$.
We proceed similarly for every pair $(i,j)$ such that $R_{ij} = R_{ji} = 1$.

\vspace{0.1cm}
Finally, we note that the symplectic transformation associated with the operator $H_N t_R H_N$ is 
$\Omega\, T_R\, \Omega = \begin{bmatrix} 
I_m & 0 \\ 
R & I_m 
\end{bmatrix}$.

\item $g_t = H_{2^t} \otimes I_{2^{m-t}} \colon$ Since $H_{2^t}$ is the $t$-fold Kronecker product of $H$ and since $D(a,b) = X^{a_1} Z^{b_1} \otimes \cdots \otimes X^{a_m} Z^{b_m}$, we have $g_t D(a,b) g_t^{\dagger}$
\begin{align}
& = \left( Z^{a_1} X^{b_1} \otimes \cdots \otimes Z^{a_t} X^{b_t} \right) \nonumber \\
  & \quad  \otimes \left( X^{a_{t+1}} Z^{b_{t+1}} \otimes \cdots \otimes X^{a_m} Z^{b_m} \right) \\
  & = \left( (-1)^{a_1 b_1} X^{b_1} Z^{a_1} \otimes \cdots \otimes (-1)^{a_t b_t} X^{b_t} Z^{a_t} \right) \nonumber \\
  & \quad  \otimes \left( X^{a_{t+1}} Z^{b_{t+1}} \otimes \cdots \otimes X^{a_m} Z^{b_m} \right).
\end{align}
We write $(a,b) = (\hat{a} \bar{a}, \hat{b} \bar{b})$, where $\hat{a} \triangleq a_1 \cdots a_t, \bar{a} \triangleq a_{t+1} \cdots a_m, \ \hat{b} \triangleq b_1 \cdots b_t, \bar{b} \triangleq b_{t+1} \cdots b_m$.
Then
\begin{align}
g_t D(\hat{a} \bar{a}, \hat{b} \bar{b}) g_t^{\dagger} & = (-1)^{\hat{a} \hat{b}^T} D(\hat{b} \bar{a}, \hat{a} \bar{b}) \\
  & = (-1)^{\hat{a} \hat{b}^T} D \left( [\hat{a} \bar{a}, \hat{b} \bar{b}] G_t \right), \\
\text{where} \ 
G_t & = 
\begin{bmatrix}
0 & 0 & I_t & 0 \\
0 & I_{m-t} & 0 & 0 \\
I_t & 0 & 0 & 0 \\
0 & 0 & 0 & I_{m-t}
\end{bmatrix}.
\end{align}
Defining $U_t \triangleq \begin{bmatrix} I_t & 0 \\ 0 & 0 \end{bmatrix}, L_{m-t} \triangleq \begin{bmatrix} 0 & 0 \\ 0 & I_{m-t} \end{bmatrix}$, we then write $G_t = 
\begin{bmatrix} 
L_{m-t} & U_t \\ 
U_t & L_{m-t} 
\end{bmatrix}$.
Similar to part 1 above, the circuit for $g_t$ is simply $H$ applied to each of the first $t$ qubits.
Although this is a special case where the Hadamard operator was applied to consecutive qubits, we note that the symplectic transformation for Hadamards applied to arbitrary non-consecutive qubits can be derived in a similar fashion.

\end{enumerate}
Hence, we have demonstrated the elementary symplectic transformations in $\text{Sp}(2m,\mathbb{F}_2)$ that are associated with arbitrary Hadamard, Phase, Controlled-$Z$ and Controlled-NOT gates.
Since we know that these gates, along with $HW_N$, generate the full Clifford group~\cite{Gottesman-arxiv09}, these elementary symplectic transformations form a universal set corresponding to physical operators in the Clifford group.

\section{Proof of Theorem~1}
\label{sec:trung_proof}

Let $F = \begin{bmatrix}
A & B \\
C & D
\end{bmatrix}$ so that $\begin{bmatrix} A & B \end{bmatrix} \Omega \begin{bmatrix} A & B \end{bmatrix}^T = 0$ and $\begin{bmatrix} C & D \end{bmatrix} \Omega \begin{bmatrix} C & D \end{bmatrix}^T = 0$ since $F \Omega F^T = \Omega$.
We will perform a sequence of row and column operations to transform $F$ into the form $\Omega \, T_{R_1} \Omega$ for some symmetric $R_1$.
If rank$(A) = k$ then there exists a row transformation $Q_{11}^{-1}$ and a column transformation $Q_2^{-1}$ such that
$Q_{11}^{-1} A Q_2^{-1} = \begin{bmatrix}
I_k & 0 \\
0 & 0
\end{bmatrix}$.
Using the notation for elementary symplectic transformations discussed above, we apply $Q_{11}^{-1}$ and $A_{Q_2^{-1}}$ to $\begin{bmatrix} A & B \end{bmatrix}$ and obtain
\begin{align*}
\begin{bmatrix} Q_{11}^{-1} A & Q_{11}^{-1} B \end{bmatrix} \begin{bmatrix} Q_2^{-1} & 0 \\ 0 & Q_2^T \end{bmatrix} & =
\left[ \begin{array}{cc|cc}
I_k & 0 & R_k & E' \\
0 & 0 & E & B_{m-k}
\end{array} \right] \\
 & \triangleq \begin{bmatrix} A' & B' \end{bmatrix} ,
\end{align*}
where $B_{m-k}$ is an $(m-k) \times (m-k)$ matrix.
Since the above result is again the top half of a symplectic matrix, we have $\begin{bmatrix} A' & B' \end{bmatrix} \Omega \begin{bmatrix} A' & B' \end{bmatrix}^T = 0$ which implies $R_k$ is symmetric, $E = 0$ and hence rank$(B_{m-k}) = m-k$.
Therefore we determine an invertible matrix $Q_{m-k}$ which transforms $B_{m-k}$ to $I_{m-k}$ under row operations.
Then we apply $Q_{12}^{-1} \triangleq \begin{bmatrix}
I_k & 0 \\
0 & Q_{m-k}
\end{bmatrix}$: 
\begin{align*}
& \begin{bmatrix} Q_{12}^{-1} Q_{11}^{-1} A & Q_{12}^{-1} Q_{11}^{-1} B \end{bmatrix} \begin{bmatrix} Q_2^{-1} & 0 \\ 0 & Q_2^T \end{bmatrix} \\
  & =
\left[ \begin{array}{cc|cc}
I_k & 0 & R_k & E' \\
0 & 0 & 0 & I_{m-k}
\end{array} \right] .
\end{align*}
Now we observe that we can apply row operations to this matrix and transform $E'$ to $0$.
We left multiply by $Q_{13}^{-1} \triangleq \begin{bmatrix}
I_k & E' \\
0 & I_{m-k}
\end{bmatrix}$:
\begin{align*}
& \begin{bmatrix} Q_{13}^{-1} Q_{12}^{-1} Q_{11}^{-1} A & Q_{13}^{-1} Q_{12}^{-1} Q_{11}^{-1} B \end{bmatrix} \begin{bmatrix} Q_2^{-1} & 0 \\ 0 & Q_2^T \end{bmatrix} \\
  & =
\left[ \begin{array}{cc|cc}
I_k & 0 & R_k & 0 \\
0 & 0 & 0 & I_{m-k}
\end{array} \right] .
\end{align*}
Since the matrix $R_2 \triangleq \begin{bmatrix}
R_k & 0 \\
0 & 0
\end{bmatrix}$ is symmetric, we apply the elementary transformation $T_{R_2}$ from the right to obtain
\begin{align*}
& \left[ \begin{array}{cc|cc}
I_k & 0 & R_k & 0 \\
0 & 0 & 0 & I_{m-k}
\end{array} \right] 
\left[ \begin{array}{cc|cc}
I_k & 0 & R_k & 0 \\
0 & I_{m-k} & 0 & 0 \\
\hline
0 & 0 & I_k & 0 \\
0 & 0 & 0 & I_{m-k}
\end{array} \right] \\
  & =
\left[ \begin{array}{cc|cc}
I_k & 0 & 0 & 0 \\
0 & 0 & 0 & I_{m-k}
\end{array} \right] .
\end{align*}
Finally, we apply the elementary transformation $G_k \Omega = \begin{bmatrix} 
U_k & L_{m-k} \\
L_{m-k} & U_k  
\end{bmatrix}$ to obtain
\begin{align*}
& \left[ \begin{array}{cc|cc}
I_k & 0 & 0 & 0 \\
0 & 0 & 0 & I_{m-k}
\end{array} \right]
\left[ \begin{array}{cc|cc}
I_k & 0 & 0 & 0 \\
0 & 0 & 0 & I_{m-k} \\
\hline
0 & 0 & I_k & 0 \\
0 & I_{m-k} & 0 & 0
\end{array} \right] \\
  & =
\left[ \begin{array}{cc|cc}
I_k & 0 & 0 & 0 \\
0 & I_{m-k} & 0 & 0
\end{array} \right]
=
\begin{bmatrix}
I_m & 0
\end{bmatrix} .
\end{align*}
Hence we have transformed $F$ to the form $\Omega \, T_{R_1} \Omega = \begin{bmatrix}
I_m & 0 \\
R_1 & I_m
\end{bmatrix}$, i.e., if we define $Q_1^{-1} \triangleq Q_{13}^{-1} Q_{12}^{-1} Q_{11}^{-1}$ then we have
\begin{align*}
A_{Q_{1}^{-1}} F A_{Q_2^{-1}} T_{R_2} G_k \Omega = \Omega \, T_{R_1} \Omega .
\end{align*} 
Rearranging terms and noting that $A_{Q}^{-1} = A_{Q^{-1}},  \Omega^{-1} = \Omega, G_k^{-1} = G_k, T_{R_2}^{-1} = T_{R_2}$ we obtain $F = A_{Q_1} \Omega \, T_{R_1} G_k T_{R_2} A_{Q_2}$.  \hfill \IEEEQEDhere


\section{{MATLAB\textsuperscript{\textregistered}} Code for Algorithm~2}
\label{sec:alg2_matlab}

\begin{lstlisting}
function F_all = find_all_symp_mat(U,V,I,J)

I = I(:)';
J = J(:)';
Ibar = setdiff(1:m,I);
Jbar = setdiff(1:m,J);
alpha = length(Ibar) + length(Jbar); 
tot = 2^(alpha*(alpha+1)/2);
F_all = cell(tot,1);

% Find one solution using symplectic transvections (Algorithm 1)
F0 = find_symp_mat(U([I, m+J], :), V);

A = mod(U * F0, 2);
Ainv = gf2matinv(A);  
IbJb = union(Ibar,Jbar);
Basis = A([IbJb, m+IbJb],:);  % these rows span the subspace W^{\perp} in Theorem 7
Subspace = mod(de2bi((0:2^(2*length(IbJb))-1)', 
                    2*length(IbJb)) * Basis, 2);

% Collect indices of free vectors in the top and bottom halves of Basis
% Note: these are now row indices of Basis, not row indices of A!!
[~, Basis_fixed_I, ~] = intersect(IbJb,I);  % intersect(IbJb,I) = intersect(I,Jbar)
[~, Basis_fixed_J, ~] = intersect(IbJb,J);  % intersect(IbJb,J) = intersect(Ibar,J)
Basis_fixed = [Basis_fixed_I, length(IbJb) + Basis_fixed_J];
Basis_free = setdiff(1:2*length(IbJb), Basis_fixed);

Choices = cell(alpha,1);

% Calculate all choices for each free vector using just conditions imposed
% by the fixed vectors in Basis (or equivalently in A)
for i = 1:alpha
    ind = Basis_free(i);
    h = zeros(1,length(Basis_fixed));    
    % Impose symplectic inner product of 1 with the "fixed" symplectic pair
    if (i <= length(Ibar))
        h(Basis_fixed == length(IbJb) + ind) = 1;
    else
        h(Basis_fixed == ind - length(IbJb)) = 1;
    end    
    % Check the necessary conditions on the symplectic inner products
    Innpdts = mod(Subspace * fftshift(Basis(Basis_fixed,:), 2)', 2);
    Choices{i,1} = Subspace(bi2de(Innpdts) == bi2de(h), :);
end

% First free vector has 2^(alpha) choices, second has 2^(alpha-1) choices and so on
for l = 0:(tot - 1)
    Bl = A;
    W = zeros(alpha,2*m);   % Rows are choices made for free vectors
    % W(i,:) corresponds to Basis(Basis_free(i),:)
    lbin = de2bi(l,alpha*(alpha+1)/2,'left-msb');
    v1_ind = bi2de(lbin(1,1:alpha),'left-msb') + 1;
    W(1,:) = Choices{1,1}(v1_ind,:);
    for i = 2:alpha
        % vi_ind loops through the 2^(alpha-(i-1)) valid choices for the i-th free vector
        vi_ind = bi2de(lbin(1,sum(alpha:-1:alpha-(i-2)) + (1:(alpha-(i-1)))),'left-msb') + 1;
        Innprods = mod(Choices{i,1} * fftshift(W,2)', 2);        
        % Impose symplectic inner product of 0 with chosen free vectors
        h = zeros(1,alpha);
        % Handle case when Basis contains a symplectic pair of free vectors
        if (i > length(Ibar))
            h(Basis_free == Basis_free(i) - length(IbJb)) = 1;
        end
        % Check the necessary and sufficient conditions on the symplectic inner products
        Ch_i = Choices{i,1}(bi2de(Innprods) == bi2de(h), :);
        W(i,:) = Ch_i(vi_ind,:);  % use the vi_ind-th valid choice for the i-th free vector
    end
    Bl([Ibar, m+Jbar], :) = W;  % replace rows of free vectors with current choices
    F = mod(Ainv * Bl, 2);          % this is the matrix F' in Theorem 7
    F_all{l+1,1} = mod(F0 * F, 2);
end    

end
\end{lstlisting}


\section*{Acknowledgment}

We would like to thank Jungsang Kim for pointing us to the work of Chao and Reichardt, and Jianfeng Lu for helpful discussions regarding Theorem~\ref{thm:symp_lineq_all}.
We are also very thankful to the anonymous reviewers (including at conferences where this work was presented as a poster) who provided helpful feedback to improve the presentation and keep the paper as self-contained as possible. 
In particular, we would like to thank one reviewer who pointed out the example in Remark~\ref{rem:stab_freedom} and suggested us to specifically mention that we are ignoring the stabilizer degrees of freedom.
We would also like to thank Jean-Pierre Tillich for highlighting that the proof of Theorem~\ref{thm:symp_lineq_all} is incomplete without the counting argument.
S. Kadhe would like to thank Robert Calderbank for his hospitality during S. Kadhe's visit to Duke University.

\IEEEtriggeratref{56}

\EOD

\end{document}